# The Effects of Stellar Rotation.

# II. A Comprehensive Set of Starburst99 Models

Claus Leitherer

*Space Telescope Science Institute[1], 3700 San Martin Drive, Baltimore, MD 21218, USA*

*leitherer@stsci.edu*

Sylvia Ekström, Georges Meynet, Daniel Schaerer

*Observatoire de Genève, 51 chemin des Maillettes, CH 1290 Versoix, Switzerland*

*sylvia.ekstrom@unige.ch, georges.meynet@unige.ch, daniel.schaerer@unige.ch*

Katerina B. Agienko

*Main Astronomical Observatory, National Academy of Sciences of Ukraine,*

*ul. Akademika Zabolotnogo 27, Kyiv, 03680, Ukraine*

*agienko@mao.kiev.ua*

Emily M. Levesque

*CASA, Department of Astrophysical and Planetary Sciences,*

*University of Colorado 389-UCB, Boulder, CO 80309, USA*

*emily.levesque@colorado.edu*

# Abstract

We present a new set of synthesis models for stellar populations obtained with Starburst99, which are based on new stellar evolutionary tracks with rotation. We discuss models with zero rotation velocity and with velocities of 40% of the break-up velocity on the zero-age main-sequence. These values are expected to bracket realistic rotation velocity distributions in stellar populations. The new rotating models for massive stars are more luminous and hotter due to a larger convective core and enhanced surface abundances. This results in pronounced changes in the integrated spectral energy distribution of a population containing massive stars. The changes are most significant at the shortest wavelengths where an increase of the ionizing luminosity by up to a factor of 5 is predicted. We also show that high equivalent widths of recombination lines may not necessarily indicate a very young age but can be achieved at ages as late as $\sim 10^7$ yr. Comparison of these two boundary cases (0 and 40% of the break-up velocity) will allow users to evaluate the effects of rotation and provide guidance for calibrating the stellar evolution models. We also introduce a new theoretical ultraviolet spectral library built from the Potsdam Wolf-Rayet (PoWR) atmospheres. Its purpose is to help identify signatures of Wolf-Rayet stars in the ultraviolet whose strength is sensitive to the particulars of the evolution models. The new models are available for solar and $1/7^{th}$ solar metallicities. A complete suite of models can be generated on the Starburst99 website (www.stsci.edu/science/starburst99/). The updated Starburst99 package can be retrieved from this website as well.

---

[1] Operated by the Association of Universities for Research in Astronomy, Inc., under NASA contract NAS5−26555

*Key words:* stars: atmospheres – stars: evolution – stars: rotation – galaxies: starburst – galaxies: stellar content

## Introduction

The evolutionary synthesis code Starburst99 calculates theoretical properties of unresolved stellar population with an emphasis on young, massive stars (Leitherer et al. 1999; Vázquez & Leitherer 2005; Leitherer & Chen 2009). In the numerical simulations, stars of an initial chemical composition form according to an initial mass function (IMF), are spread out along the zero-age main-sequence, evolve across the Hertzsprung-Russell (HR) diagram, and reach their final states as black holes, neutron stars/pulsars, or white dwarfs. At each point in the stellar life and for each mass we assign spectra or other desired quantities and then sum over all stellar masses and generations in order to obtain the integrated population properties. Conroy (2013) provided a comprehensive review of the synthesis technique and its challenges. Stellar evolution models are at the heart of any population synthesis code. Stellar evolution provides the relation between stellar mass ($M$) and luminosity ($L$) and their evolution with time. Any uncertainties, revisions, or errors in stellar evolution models will immediately affect the predictions of synthesis codes. With respect to populations of massive stars resulting from a given star-formation history and a given IMF, the uncertainties of stellar evolutionary tracks are arguably the single most important concern for the computed synthetic properties.

Synthesis of massive star populations came of age with the advent of stellar evolution models accounting for mass loss (Chiosi & Maeder 1986). Massive stars of any temperature develop strong stellar winds and/or eruptions which lead to a significant decrease of *M* over the stellar life-time and therefore are a necessary condition for understanding the evolutionary connections in the upper HR diagram. A classic example is the "Conti Scenario" (Conti 1976) which explains the evolution of massive O stars into low-mass Wolf-Rayet (W-R) stars as the consequence of strong mass loss by stellar winds. The concept of mass loss has been incorporated in all stellar evolution models. The evolutionary tracks for massive stars most widely used in population synthesis are those published by the Geneva (Schaller et al. 1992; Schaerer et al. 1993a; 1993b; Charbonnel et al. 1993; Meynet et al. 1994) and the Padova (Bressan et al.1993; Fagotto et al. 1994a; 1994b; Girardi et al. 2000) groups. These tracks are currently implemented in Starburst99 as well.

At the time of publication of the above evolution models the general thought was that stellar mass loss is the single most – if not the dominant – driver of massive star evolution. Since then, two key developments have occurred. First, the mass-loss rates ($\dot{M}$) by stellar winds of hot stars had been overestimated by factors of several because wind inhomogeneities had not been taken into account (Puls, Vink, & Najarro 2008). This by itself would not invalidate the Conti scenario because mass loss can still occur in short eruptive phases during the Luminous Blue Variable phase (Smith 2009). If these phases are more important than previously assumed, they can compensate for the reduced mass loss in the stationary wind phase. However, eruptive mass loss cannot account for the second development: stars which should not have strong stellar winds during their main-sequence phase (typically 15 M$_\odot$ stars) and thus should not display any changes of their surface abundances, may actually show nitrogen surface enrichments (Hunter et al.

2009). Since nitrogen is the tell-tale sign of nuclear burning, a mechanism to expose the central material at the surface is needed. In the absence of removal of surface material by winds, this requires an effective mixing process. This process is thought to be rotation.

Maeder & Meynet (2000; 2012) reviewed the fundamental physical processes associated with massive star rotation. Since publication of their article, the Geneva group released multiple generations of stellar evolution models with rotation. Many of these models were exploratory and/or covered a limited parameter space, which makes them less suitable for implementation. Nevertheless, Vázquez et al. (2007) performed an initial analysis of the impact of stellar rotation on the predictions of Starburst99. Vázquez et al. found quite dramatic consequences. Since then, the Geneva group generated complete sets of evolutionary tracks for solar and subsolar chemical composition in a format compatible for population synthesis (Ekstrom et al. 2012 [E12]; Georgy et al. 2013 [G13]). Chieffi & Limongi (2013) also released models for the pre-supernova evolution of rotating solar metallicity stars. Their grid is restricted to the mass range $13 - 120$ $M_\odot$, as opposed to the full mass range down 0.8 $M_\odot$ in the Geneva models. This limitation makes them less useful for implementation in our synthesis models. An initial demonstration of the impact of the Geneva evolutionary tracks on the predictions of Starburst 99 was given by Levesque et al. (2012; Paper I) who discussed the revisions to the ionizing spectra. In the present paper we fully describe the implementation of the tracks in Starburst99 (Section 2). Since W-R stars are of particular interest for testing model predictions, we optimized our ultraviolet (UV) spectral library by adding a grid of W-R spectra. This effort is described in Section 3. A comparison of the model predictions with different stellar evolution models is presented in Section 4. In Section 5 we provide a discussion of the results. The conclusions are in Section 6.

# 1. Implementation of the stellar evolution models in Starburst99

Two sets of evolutionary tracks with rotation have been released by the Geneva group: one set of tracks for solar chemical composition (E12), and another for subsolar composition (G13). Throughout this paper we will refer to the former set as "v40-h" and to the latter as "v40-l". Both sets of models assume an initial rotation velocity $v_{rot}$ of 40% of the break-up velocity $v_{break}$ on the zero-age main-sequence. E12 and G13 also published sets of tracks with zero rotation velocity but with otherwise the same parameters as the v40-h and v40-l models. We refer to these models as "v00-h" and "v00-l". The non-rotating models serve as an important benchmark for isolating the effects of rotation on synthetic population properties when making comparisons with earlier evolution models, which often differ in additional parameters, such as chemical composition or mass-loss rates. The previously recommended default set of Geneva tracks in Starburst99 are the models published by Meynet et al. (1994 [M94]). These are often referred to as the high mass loss models since $\dot{M}$ was increased by a factor of two during all the stellar evolutionary phases except the W-R phase for better agreement with observations. Again we selected sets with solar ("1994-h") and subsolar ("1994-l") chemical composition and included them in our study of the effects of rotation on the population properties. Table 1 summarizes the most relevant properties of the six sets of evolutionary tracks.

The initial heavy-element abundance $Z$ (by mass) is in column 3 of Table 1. The v40-h and v00-h models adopted the revised solar abundances of Asplund et al. (2005), which have oxygen at log (O/H) + 12 = 8.66. The corresponding heavy-element abundance is $Z = 0.014$. The v40-l and v00-l models use $Z = 0.002$, i.e., 1/7 the value of the solar models. The element ratios in the metal-poor models are the same as in the solar models. The M94 tracks are in the old solar composition reference system, i.e., $Z = 0.020$ for the 1994-h models. Our selected 1994-l model has $Z = 0.004$, which is twice the value of the v40-l and v00-l tracks. The set of evolution models published by M94 only includes abundances of $Z = 0.040, 0.020, 0.008, 0.004$, and $0.001$. We opted for the 0.004 tracks for our comparison.

Column 4 of Table 1 lists the rotation velocities of the tracks. The ratio of the length of the convective overshooting $d_{over}$ relative to the pressure scale height $H_P$ is in column 5. This parameter cannot be derived from first principles and is usually calibrated by the width of the main-sequence in the HR diagram. Observations suggest $d_{over}/H_P = 0.1 - 0.2$. The older 1994-h and 1994-l models use a value of 0.2, whereas the latest models prefer 0.1 since part of the enlargement of the core which had previously completely been attributed to the effect of overshooting is also due, in part, to rotational mixing in the rotating models. Mass loss by stationary winds occurs in three major evolutionary phases: during core-hydrogen burning in the OB phase, during core-hydrogen or -helium burning in the hot part of the HR diagram in the W-R phase, and during core-helium burning in the cool red supergiant (RSG) phase. The evolution models are very sensitive to the rates in each of the phases. We quote the adopted prescriptions $\dot{M}_\text{O}$, $\dot{M}_\text{W-R}$, and $\dot{M}_\text{RSG}$ in columns 6, 7, and 8 of Table 1. The v40-h and v00-h tracks adopt the theoretical rates of Vink, de Koter, & Lamers (2001) for $\dot{M}_\text{O}$. These rates are lower by factors of several than the observed rates, as expected since the observed rates are affected by inhomogeneities. $\dot{M}_\text{W-R}$ is based on empirical values from Nugis & Lamers (2000) and Gräfener & Hamann (2008), which are already downward corrected for inhomogeneities. For the RSG phase we used the Reimers (1975; 1977) formula for stars up to 12 $M_\odot$ and de Jager et al. (1988) for RSGs with masses of 15 $M_\odot$ and above. In addition, the RSG rates are increased by a factor of 3 whenever the Eddington luminosity is exceeded. The v40-l and v00-l tracks use the same prescriptions as their metal-rich counterparts, but with an additional scaling for the $Z$ dependence. The scaling factors are given in Table 1. Both O- and W-R stars have a mild metallicity dependence (see also the discussion in Section 3) whereas no such dependence is assumed for RSGs. The 1994-h evolution models adopt the de Jager, Nieuwenhuijzen, & van der Hucht (1988) mass loss parameterization in the entire HR diagram, scaled by a factor of 2, except for the W-R phase. During the latter, $\dot{M} = 8\times10^{-5}$ $M_\odot$ yr$^{-1}$ is assumed for WNL stars, and the relation of Langer (1989) is used for all other W-R types. The metal-poor models apply scaling factors to the rates of 1994-h. Note that the 1994-l models assume no metallicity dependence for W-R winds but the same dependence for RSGs as for O stars, which is the opposite of what is assumed in the v40-l and v00-l models. The major difference between the 1994 and v40/v00 models at any chemical composition are lower rates during hot stellar phases and higher rates in the RSG phase for the latter.

The integration of the evolutionary tracks into Starburst99 follows the same scheme as in our previous work, and the reader is referred to Leitherer et al. (1999) for details. An important aspect of evolutionary models of hot, massive stars is the definition of temperature. The Geneva models distinguish between the hot, hydrostatic core temperature $T_*$ and the cooler surface temperature $T_{2/3}$. We call the latter $T_{eff}$. $T_{2/3}$ refers to the stellar radius where optical depth 2/3 is reached. Depending on the wind density, this radius can be substantially larger than the hydrostatic radius and consequently $T_{2/3}$ can be lower than $T_*$. Starburst99 uses the extended, expanding model atmospheres WM-Basic (Pauldrach et al. 1998), CMFGEN (Hillier & Miller 1998, 1999), and PoWR (Gräfener, Koesterke, & Hamann 2002; Hamann & Gräfener 2003, 2004) for its various outputs. These atmospheres are defined by $T_*$ and need to be assigned to the evolution models with an appropriate description. As we did before, we follow the empirical prescription of Smith et al. (2002) who adopted a hybrid weighting of 0.6 $T_*$ and 0.4 $T_{2/3}$ to match the tracks and the atmospheres. In Section 5 we will discuss how this assumption affects the outcome. The chemical composition of the atmospheres available in Starburst is tied to the 1994-h and 1994-l models. Therefore we link the v40-h and v00-h evolution models to atmospheres with $Z = 0.020$ and the v40-l and v00-l models to atmospheres having $Z = 0.004$.

We use definitions for hot stars which are similar to those used in the evolutionary tracks. O stars are counted by utilizing the spectral-type versus $T_{eff}$ calibration of Martins, Schaerer, & Hillier (2005). Stars with hydrogen surface abundances of less than 0.4 are excluded since these stars are considered W-R stars. We perform a nearest-neighbor search in $T_{eff}$ and $L$, assign spectral types, and sum over all spectral types. The total number is defined to be the number of O stars. In practice, this procedure is equivalent to defining O stars as having $T_{\rm eff} > 30,000$ K. W-R stars must meet three conditions: log $T_{\rm eff}$ > 4.4, hydrogen surface abundance < 0.4, and flagged by the evolutionary models to originate from an initial mass capable for forming W-R stars. After the stars have passed the initial W-R screening, we distinguish between WN and WC stars using the surface abundances of helium, carbon, and nitrogen. If the surface hydrogen abundance is > 0.1, it is a WNL star, if it is less and the C/N abundance ratio is < 10, it is a WNE star. Otherwise it is classified as a WC star. Again, we will discuss how these definitions affect the W-R star counts in Section 5.

## 2. The Potsdam Wolf-Rayet star library

Starburst99 allows users to generate synthetic spectra in the satellite-UV between 912 and 3000 Å at 0.4 Å resolution. The spectra utilize a theoretical spectral library of OB stars computed with the radiation-hydrodynamics code WM-Basic (Leitherer et al. 2010). This library does not include W-R stars. In order to account for the contribution of W-R stars to the UV spectral region, the low-resolution (10 Å) W-R library of Smith, Norris, & Crowther (2002) is added to the OB library. This approach accounts for the contribution of W-R stars to the continuum and some of the strongest lines but does not permit accurate line diagnostics, should W-R stars become significant in the synthetic population spectrum. Given the heightened importance of hot stars in the new evolution models, we were motived to complement the existing OB library with W-R spectra of comparable quality.

The Potsdam Wolf-Rayet (PoWR) models are a comprehensive grid of W-R spectra of the WN and WC sequence available for download at a designated website[2]. The PoWR code computes non-LTE atmospheres with complex model atoms including iron-line blanketing. One important difference between O- and W-R winds are the higher wind densities of the latter. Hot-star winds are driven by momentum transfer from photons to metal ions. The high W-R wind densities are thought to result from multiple photon scatterings due to the high *L/M* ratio in W-R stars. The PoWR and the CMFGEN model atmosphere codes are the two most widely used tools for generating W-R spectra. Since the PoWR grid is readily available and well-suited for the construction of a stellar library, we opted for these models.

The three major parameters describing W-R spectra are *L*, $T_*$, and the wind density. Schmutz, Hamann, & Wessolowski (1989) demonstrated that the wind density can be parameterized to define a "transformed" radius

$$R_T = R_* \left( \frac{v_\infty}{2500} \Big/ \frac{\sqrt{D}\,\dot{M}}{10^{-4}} \right)^{2/3}. \qquad (1)$$

In this equation $R_*$ is the stellar radius, $v_\infty$ is the terminal wind velocity in km s$^{-1}$, $\dot{M}$ is the mass-loss rate in M$_\odot$yr$^{-1}$, and *D* is the "clumping factor", which accounts for wind inhomogeneities. Models with identical values of $R_T$ have rather similar normalized line spectra. For a pre-specified luminosity, the W-R spectrum will only depend on $T_*$ and $R_T$. We defined the parameter space of the W-R library guided by $T_*$ of the evolutionary tracks in the HR diagram. The resulting parameter set closely mirrors the parameters adopted by Smith et al. (2002) for their low-resolution CMFGEN library. We selected 12 models for both the WN and the WC sequence whose properties are given in Table 2.

---

[2] http://www.astro.physik.uni-potsdam.de/~wrh/PoWR/powrgrid1.html

Column 1 of this table lists the sequence number. Note that these are *not* W-R subtypes. $T_*$ is in column 2. The temperature is the equivalent to the hotter core temperature in the Geneva evolutionary tracks. $R_*$ in column 3 was obtained from $T_*$ and log $L$, which is normalized at 5.3 in all models. The stellar-wind parameters $\dot{M}$ (column 4), $D$ (column 5), and $v_\infty$ (column 6) are from Crowther (2007). Crowther gives "clumping-free" mass-loss rates of log $\dot{M} = -5.0$ and $-4.8$ for WN and WC stars, respectively. The clumping factors are 4 and 10. We assumed that WN models with log $T_*$ between 4.5 and 4.65 correspond to WNL stars, whereas hotter models are of type WNE. The former still have residual hydrogen in their atmospheres (20% by mass in the PoWR models), whereas the latter are H-free. According to Crowther, WNL stars have $v_\infty = 1000$ km s$^{-1}$. WNE stars, as well as all WC types, have $v_\infty = 2000$ km s$^{-1}$. $R_T$ obtained with equation (1) is in column (7). We then picked the appropriate PoWR model for each pair of $T_*$ and $R_T$. For easy identification, we give the corresponding PoWR designation in column 8 of Table 2.

    The PoWR models discussed here are for solar initial chemical composition, with subsequent modification of (primarily) H, He, C, N, and O. Since the chemical composition of W-R stars is determined by interior nucleosynthesis, one would expect no dependence of the W-R spectra on the initial chemical composition. In reality, there is observational evidence of at least some metallicity dependence of the W-R properties (Crowther 2006; Crowther & Hadfield 2006). Since the W-R spectra are determined by the wind properties, metallicity dependent winds will modify the spectra. If the winds were driven by CNO elements, no such dependence would be expected. However, wind models suggest that Fe-peak elements are the main driving agents, and Fe does of course vary with the environment. The resulting effect is subtle but needs study in the future.

Currently there are no very metal-poor PoWR spectra available, and we used the solar models for all simulations presented in this paper.

We implemented the PoWR spectra in Starburst99 following the prescriptions discussed in Section 2. After the stars enter the W-R phase, they are first classified as WNL, WNE, or WC. Then a nearest-neighbor search in $T_*$ is performed and a spectrum is assigned. Together with all other OB stars, this spectrum is then used to build the full population spectrum. Except for the newly added PoWR spectra, which are replacing the low-resolution CMFGEN spectra, the models for the 1994-h and 1994-l tracks presented in the following are identical to those in Leitherer et al. (2010).

We are plotting the evolution of the spectrum between 1200 and 2500 Å for a single stellar population with ages from 1 to 9 Myr in Figure 1 and Figure 2 using the 1994-h and 1994-l tracks, respectively. In the case of solar chemical composition, W-R spectral features introduced by the new PoWR library are conspicuous between 3 and 6 Myr. The location of some of the strongest lines occurring in individual W-R stars are indicated by vertical lines. Not all of the features are actually present in the series of population spectra shown in Figure 1.The strongest W-R features in Figure 1 are He II 1640 and N III 2064. Relatively few observations of the spectral region around 2000 Å in star-forming galaxies exist. The reasons are the lack of diagnostic lines in this region and the location of this wavelength range at either the extreme long end of a far-UV detector or the short end of a near-UV detector. Nevertheless, there are indications for the presence of these W-R features in some galaxies. Inspection of Figure 4 of the spectral atlas of Leitherer et al. (2011) suggests that N III 2064 is present in I Zw 18, NGC 2403, and NGC 2366. The W-R features are even stronger in the two other metal-rich models (v00-h, see Figure 3; v40-h, see Figure 5) but not at low metallicity (1994-l, see Figure 2; v00-l, see Figure 4; v40-l, see Figure 6). The absence simply results from the low W-R numbers predicted in the metal-poor models. We are not performing a comparison between the different evolutionary models, which will be done in the following section. Rather, the purpose of Figure 1 through Figure 6 is to demonstrate that irrespective of the choice of the evolution models, we have a library in place which produces W-R features, should they be predicted by the evolution models.

## 3. Parameter study

We generated a grid of model predictions with Starburst99 in order to highlight any revisions to the population properties caused by the new stellar evolution models. This is not supposed to be a discussion of the full suite of predictions which can be generated with Starburst99. Rather, the examples are intended to identify the main differences and connect them with their physical origin in the evolution models. Throughout this section we assume a single stellar population (SSP) with a mass of $10^6$ M$_\odot$ and following a Kroupa (2008) IMF between 0.1 and 100 M$_\odot$. This IMF is a two-component power-law with slopes of 1.3 and 2.3 below and above 0.5 M$_\odot$, respectively. In the high-mass regime, this IMF is essentially a Salpeter IMF. All models shown here extend to ages of $10^8$ yr, corresponding to the lifetime of a ~6 M$_\odot$ star.

The extreme-UV to near-IR spectral energy distributions (SEDs) for the v40-h, v00-h, and 1994-h tracks are shown in Figure 7. The resolution is 10 – 20 Å, depending on wavelength. We provide age snapshots at 4, 10, and 20 Myr. The general trend is a harder spectrum of the rotating models in the UV and relatively minor differences between the 1994-h and v00-h models. The models with rotation have longer lifetimes by 10 – 20%. Therefore some of the differences simply result from observing a slightly different evolutionary phase at a given age in the three sets of tracks. For instance, the bulge in the SED between 1 and 2 μm at ages 10 and 20 Myr indicates RSGs. Comparing the SEDs with the behavior of the CO index (Figure 22, to be discussed below) explains the behavior of the near-IR index in the different evolutionary tracks. The corresponding SEDs for the v40-l, v00-l, and 1994-l tracks are in Figure 8. The trends are similar to those at higher metallicity. The most notable difference is the weakness of the near-IR bulge due to the lower number of RSGs at lower metallicity. Optical high-resolution (0.3 Å) spectra between 3500 and 6000 Å for the v40-h, v00-h, and 1994-h tracks are reproduced in Figure 9. As before we provide age snapshots at 4, 10, and 20 Myr. The most conspicuous features are the W-R emission line blends of N III 4640 + C III 4650 + He II 4686 and C IV 5808. These lines are detectable at 4 Myr in all three models and are strongest in the v40-h model. This is the result of the larger number of W-R stars in the metal-rich tracks with rotation (cf. the W-R features in the UV spectra in Figure 5). At low metallicity (Figure 10), there are fewer obvious differences; all three spectra behave very similarly at all three ages. The UV spectra from 900 to 2200 Å at 0.4 Å resolution are reproduced in Figure 11 for v40-h, v00-h, and 1994-h. The age is 4 Myr when W-R features are at their peak. The 1994-h and v00-h spectra are quite similar at all wavelengths, whereas the v40-h model predicts strong W-R emission. In contrast, at low metallicity (Figure 12), only the 1994-l tracks have W-R stars in sufficient numbers to generate weak W-R emission features in the UV.

There is little difference in the supernova rate predicted by the different models, both at high and low metallicity (Figure 13). However, the progenitors of core-collapse supernovae are quite different because of the drastically different relative numbers of W-R stars, blue supergiants and RSGs. On the other hand, the ratio of W-R over O stars changes dramatically (Figure 14). The v40-h tracks generate the highest ratio at any epoch. The rotating models show a pronounced double-peak structure in the upper part of Figure 14. The second peak results from W-R stars evolving from RSGs in the ~25 $M_\odot$ range. Note that the height of the peak at $10^7$ yr (log $t$ = 7.0) is somewhat deceptive because there are few O stars left at that age and the ratio becomes degenerate. At subsolar chemical composition, only the 1994-l models predict significant W-R numbers. In Figure 15 we compare the ratio of WC over WN stars. The 1994-h models generate the highest proportion of WC stars. In both the v00-h and the v40-h models the ratio is significantly lower. In the case of v40-h the peak of WC production occurs at older ages since most WCs are produced by less massive progenitors. No WC stars are expected for the v00-l and v40-l models (lower part of Figure 15).

Models with rotation lead to a larger convective core, and therefore to larger $L$. This can be seen in Figure 16, which shows the bolometric luminosity $M_{Bol}$ of the six models. (For easy reference: log $(L/L_\odot)$ = −0.4 $M_{Bol}$ + 1.90.) Around 4 Myr (log t = 6.6), the v40-h models are ~0.5 mag more luminous than their non-rotating counterparts. The luminosity increase is slightly less pronounced for the v40-l models. This has obvious consequences, e.g., for IMF determinations where the predictions of Figure 16 are used to convert from the observed luminosity to a stellar mass. Using models with rotation will lead to a somewhat steeper IMF since the same luminosity can be reproduced by a lower mass. The graphs for the visual luminosity $M_V$ (Figure 17) exhibit no clear differences between the different tracks at both high and low metallicity. This, however, changes dramatically when moving to shorter wavelengths. In Figure 18 we show the ionizing continuum below 912 Å, expressed in terms of the number of photons capable of ionizing hydrogen. The v40-h models are both hotter (due to the larger helium surface abundance) and more luminous. The two effects combined lead to a ~5× higher photon output for the v40-h tracks when massive O stars dominate. The output for the v40-l is increased as well but to a lesser degree. If the ionizing photons are used as an indicator of current star formation, the derived rates would be lower by a corresponding factor with the rotating models. As expected, the increased UV output of the v40-h and v40-l models is even more pronounced in the neutral helium continuum shortward of 504 Å (Figure 19). This spectral range is significant for powering many important diagnostic high-ionization lines in emission-line spectra of H II regions. Models with rotation predict a much harder spectrum.

As examples of the evolution of broadband colors we are including (*B*–*V*) and (*V*–*K*) in Figure 20 and Figure 21, respectively. The various tracks predict differences of up to ~0.2 mag in (*B*–*V*), but with no clear trend. (*V*–*K*) is dominated by RSGs. At high metallicity, the v40-h models have the RSG phase at a later epoch, as indicated by the shift of the color minimum to older ages. At low metallicity, (*V*–*K*) tends to be bluer in the models with rotation due to smaller RSG numbers. These trends are mirrored by the CO index in Figure 22, which is a direct tracer of RSGs. The CO index was introduced by Doyon, Joseph, & Wright (1994) who define the index as

$$\mathrm{CO} = -2.5 \log \langle R_{2.36} \rangle, \qquad (2)$$

where $\langle R_{2.36}\rangle$ is the average of the normalized spectrum between 2.31 and 2.40 μm. Note the time delay in the v40-h models and the generally weaker CO index for both rotating models.

The equivalent width of nebular Hα emission is reproduced in Figure 23. The corresponding relations for other hydrogen recombination lines were discussed by Levesque & Leitherer (2013). The behavior of the Hα equivalent width closely mirrors the trends for the number of ionizing photons in Figure 18. We point out the important contribution of W-R stars to the equivalent width for the v40-h models. This is part of the reason for the difference between the v40-h and the v00-h/1994-h in the upper part of the figure. The 40-l models have very few W-R stars; therefore the three curves in the lower part of Figure 23 are rather similar. The presence of W-R stars can be gauged with the help of Figure 24 and Figure 25, which reproduce the equivalent widths of He II 4686 and C IV 5808, respectively. Both lines originate from W-R stars of various subtypes, and the equivalent widths plotted are essentially the ratio of the W-R stars over the continuum producing B and A stars. He II 4686 is predominantly from WN stars, which are present both in the solar and subsolar models (for the latter at greatly reduced numbers). The v40-h models predict He II 4686 up to ages of ~$10^7$ yr since significant numbers of W-R stars form from stars with progenitor masses as low as 20 $M_\odot$. C IV 5808 is mainly attributable to WC stars. The v40-h models have fewer WC stars than their non-rotating cousins, and the bulk appears at older ages. No WC stars are predicted for either the v00-l or the v40-l models (Figure 25).

The final Starburst99 predictions addressed in this paper concern non-radiative properties. In Figure 26 we show the mass loss by stellar winds and core-collapse supernovae. The values of $\dot{M}$ are from the published tables of the evolutionary tracks. When stellar winds dominate up to ~3 Myr ($\log t = 6.5$), the v00-h, v00-l, v40-h, and v40-l models are lower than the 1994-h and 1994-l models. This is mainly the result of the reduced mass-loss rates due to wind inhomogeneities. Note, however, that this effect is partly compensated by the fact that the 1994-h/l models have lower $L$. Since $\dot{M}$ scales with $L$, the rates are higher for a given initial mass in the more recent tracks. At older ages supernovae dominate, and there is little difference between the individual models. The mechanical wind luminosity for stars and supernovae is in Figure 27. The model predictions in this figure essentially are the results of the previous figure for $\dot{M}$, multiplied by $v_\infty^2$. We used the prescription of Leitherer et al. (1999) for assigning $v_\infty$.

## 4. Discussion

The new generation of stellar evolution models is intended to supersede the previous set of evolutionary models for massive stars by the Geneva group. As is the case with every release of a new model generation, some additional fine-tuning will be required. Clearly some amount of rotation is present in massive stars; therefore the v00-h/v00-l models should be considered lower limits. On the other hand, the v40-h/v40-l models have rotation velocities which may be too extreme. Martins & Palacios (2013) compared different sets of stellar evolution models for massive stars with and without rotation and found that the v40-h tracks do not reproduce the distribution of the most massive, evolved stars very well. These tracks turn to the blue side of the HR diagram already during the main-sequence evolution and do not overlap with the observed population of WN stars. This is most likely the result of too high rotation velocities, although convective overshooting and mass-loss rates may contribute as well. Therefore the v40-h/v40-l may be viewed as the upper boundary to the real case, and the tracks sets with zero and high velocities bracket the expected properties of observed populations. The diagnostic diagrams provided in this paper (and others that can be generated on the Starburst99 website) should help constraining the rotation velocities. As a first, crude approximation, users can superpose weighted fractions of v40 and v00 models to simulate more realistic rotation velocity distributions (see Fig. 8 of Paper I).

The Starburst99 code uses $L$, $T_{\rm eff}$, $T_*$, $\dot{M}$, and the surface abundances of H, He, C, N, and O as provided by the evolution models as input for the synthesis calculations. As discussed in Section 2, the use of extended model atmospheres requires a careful calibration of the weighting of $T_{\rm eff}$ and $T_*$. In its default mode, Starburst99 adopts weightings of 0.6 $T_*$ and 0.4 $T_{\rm eff}$ for connecting the atmospheres with the evolution models. In Paper I we investigated how different weighting factors affect the SEDs. $T_*$ and $T_{\rm eff}$ are almost identical during most of the stars' lifetimes and only differ when the mass-loss rates are large. This occurs in the W-R phase and for very luminous O supergiants. Therefore the effects of different weighting factors are significant in the neutral and ionized helium continuum (i.e., at wavelengths below 504 Å) where these stars dominate, but minor at longer wavelengths. In Paper I we also addressed the influence of the W-R definition (both in terms of $T_{\rm eff}$ and surface composition) on the SEDs and found that these parameters have negligible effects, except in the ionized helium continuum below 228 Å.

All Starburst99 models involving specific stellar types and spectral classes, such as W-R numbers, rely on a set of definitions and rules. The definitions adopted in Starburst99 were chosen to closely mirror those used in the stellar evolution models. The related predictions are in fact quite sensitive to these definitions. For instance, the W-R features in the UV spectra are the result of the PoWR library. Had we chosen a more restrictive W-R definition, e.g., by requiring a lower surface hydrogen abundance or higher minimum temperature to be classified as a W-R star, we would assign an O- instead of a W-R spectrum, and the features would be weakened or disappear altogether. Ultimately, the goal is to generate unified evolution models and atmospheres, which avoids the uncertainties introduced by interface issues and definition inconsistencies. A first step in this direction was made by Groh et al. (2014) who discussed the evolution of the spectrum of a 60 $M_\odot$ star from the main-sequence to the W-R phase. The spectrum was obtained by combining the Geneva evolution models and the CMFGEN atmospheres self-consistently.

Starburst99 does not produce the full nebular emission-line spectrum, except for some important recombination lines. Rather, the computed SEDs can be used as input to any of the commonly used photo-ionization codes. This has been done before using Cloudy (e.g., Wofford 2009), Mappings (e.g., Levesque & Richardson 2014), or Photo (e.g., Stasińska, Schaerer, & Leitherer 2001). Since the most pronounced effects of rotation are found in the extreme UV, the nebular spectrum can provide crucial model constraints. It is worth highlighting the behavior of the v40-h and v40-l models. A significant portion of the ionizing flux in the extreme UV is provided by W-R stars, which are abundant at solar chemical composition but nearly absent in the subsolar models. Therefore SEDs generated with the v40-h models produce harder ionizing spectra than their metal-poor counterparts. Kewley et al. (2001) and Levesque, Kewley, & Larson (2010) found a deficit of hard ionizing photons in the spectra produced by previous generations of stellar population synthesis models leading to deficiencies in the predicted [S II] fluxes. The v40-h models have the potential for generating better agreement with the observations. The mass range of stars responsible for the harder UV spectra is quite broad – it is not only stars with the highest progenitor masses but also stars evolving from the red supergiant phase back to the blue part of the HR diagram. The initial masses of such stars are as low as 20 $M_\odot$, corresponding to lifetimes of about $10^7$ yr. This is a contributing factor to the excess of the ionizing radiation of the v40-h with respect to the v00-h models between 3 and 10 Myr ($\log t = 6.5 - 7$) in Figure 18. An implication is that high equivalent widths of recombination lines may not necessarily indicate a very young age but can be achieved at ages as late as $\sim 10^7$ yr.

All stellar evolution models discussed in this paper are for single-star evolution. Recent work has suggested that at least 50% of all massive stars are in binaries and that close to 70% of massive binaries are interacting in the course of their evolution (Langer 2012; Sana et al. 2012; 2013). Processes that occur during the evolution of binaries include envelope stripping from the binary, accretion of mass by the secondary or even complete mergers (de Mink et al. 2014). Clearly the effects of binary evolution need to be considered. Georgy et al. (2012) estimated that approximately 40% of all W-R stars form via the binary channel at solar chemical composition. This fraction is almost certainly higher at subsolar composition since the mass-loss rates are lower whereas the binary fraction is about the same. Population synthesis models including the effects of binary evolution exist (Vanbeveren, Van Bever, & Belkus 2007; Eldridge 2012). However, such models have additional degrees of freedom which make them harder to constrain. While it is clear that single-star models are incomplete, they serve as the important first step of helping constrain the evolution of single stars before introducing additional complexities via binary interaction.

## 5. Conclusions

In this work we are presenting a suite of population synthesis models which incorporate new stellar evolutionary tracks for massive stars with rotation. The general trend for rotating massive stars is towards higher $L$ and hotter $T_{\text{eff}}$. This had been known from earlier work for selected mass ranges but is now presented in a complete, self-consistent way. The predictions of stellar evolution models have been compared to observations of individual stars and resolved stellar populations in the Local Group of galaxies before. Here we choose the complementary approach of studying the resulting integrated properties of unresolved populations, which can shed light on issues not easily detectable on a star by star basis (such as, e.g., the ionizing photon count). Both approaches have merits, and taken together, they can provide guidance for evolution models and help understand uncertainties when interpreting observations.

When evolution models with rotation models are chosen, the resulting SEDs of SSPs in the optical to UV become bluer and emit a harder ionizing spectrum when hot, massive stars are present. This has important consequences when using the predicted SEDs for determining stellar ages and star-formation rates. The most widely used star-formation tracers are nebular recombination lines (which are a proxy for the ionizing luminosity), the UV flux and its associated line spectrum, and the far-infrared (which is reprocessed stellar UV light). The models with rotation will lead to older ages (as the same luminosity can be reached with less massive stars), higher reddening (as the predicted SED is bluer), lower star cluster masses (as the theoretical $L/M$ is higher), a steeper IMF (as fewer massive stars are needed to account for the observed luminosity), or a combination of all these parameters. In the work, as well as in the Starburst99 code in general, we include two sets of models, one with zero rotation velocity and the other with a rotation velocity corresponding to 40% of the break-up velocity of a star on the zero-age main-sequence. These two values are considered the boundary cases and should bracket the actual situation in the observations. This approach is conceptually similar to that taken by the 1992/1994 release of Geneva evolution models. These models assumed two different values for the mass-loss rates, one called "standard" and the other called "high", with observational support in favor and against either of the two.

The major issues for massive star evolution currently under discussion are the effects of rotation, mass loss, additional mixing mechanisms, binarity, and their interplay. Constraining these properties is far from trivial. Surprisingly little is known about rotation velocities of massive stars and their variation with environment but observational data are beginning to accumulate. Large samples of Galactic OB stars were studied by Simón-Díaz & Herrero (2014) and Markova et al. (2014). Bragança et al. (2012) determined rotation velocities of 350 B stars in the Galaxy. The rotation velocities of O (Ramírez-Agudelo et al. 2013) and B stars (Dufton et al. 2013) in the 30 Doradus region in the Large Magellanic Cloud exhibit a bimodal distribution with a high-velocity tail whose origin is not understood. The fast rotators may be the result of binary interaction.

The v40-h and v40-l models assume no systematic differences of the rotation velocities at low and high metallicity other than the implicit increase of $v_{\rm rot}$ at lower metallicity due to the higher break-up velocities, which results from the shift of the main-sequence towards higher $T_{\rm eff}$. Hunter et al. (2008) found that stars in the Small Magellanic Cloud on average rotate faster than their Galactic counterparts (mainly field objects). No difference was found between Galactic stars and stars in the Large Magellanic Cloud. There is also indirect observational evidence of higher $v_{\rm rot}$ in the Small Magellanic Cloud from the larger number of Be stars, which are associated with fast rotators (Grebel, Roberts, & Brandner 1996). Additional environmental influences may be introduced by the density of the star-forming region. Strom, Wolff, & Dror (2005) found systematically larger rotation velocities of B main-sequence stars in the open clusters h and χ Persei in comparison with field stars. This result was interpreted as due to higher accretion rates in stars forming in clusters. Whereas the existence of systematic differences between rotation velocities inside and outside of clusters is fairly well established (see also Wolff et al. 2007), its interpretation is not. Huang & Giess (2008) suggested age differences as the reason for the lower rotation velocities of their sample of field B stars. If the properties of the environment were indeed relevant, this would introduce additional free parameters in the synthesis modeling.

The mass-loss rates of hot stars due to stationary winds in the new evolution models are markedly lower than those in the 1992/1994 set. This reflects the realization of an overestimate of the previous rates. The lower mass loss early in the evolution is compensated by the increased importance of the Luminous Blue Variable phase when significant amounts of mass are ejected in eruptions (see the review by Smith 2014). The differences between the v00-h/l and the 1994-h/l models are largely due to the different assumptions on $\dot{M}$. Since $\dot{M}$ scales with metallicity, the rates are much lower in the v40-l and v00-l models than in their metal-rich counterparts. Because of the low rates in metal-poor environments, very few W-R stars are predicted in the models, which may indicate a deficiency when compared with data. This serves as a reminder that all the models discussed in this work are restricted to the evolution of single stars. These models are a first step towards understanding the fundamental physics of stellar rotation in massive stars before adding an additional degree of complexity via binary evolution. Nevertheless, accounting for the evolution of both single and binary stars is the ultimate quest for population synthesis modeling.

Starburst99 v7.0.0 includes the evolution models and libraries discussed in this paper and is accessible at http://www.stsci.edu/science/starburst99/. The website provides three methods to access the models: by calculating models on-line, by accessing the fortran source code and the auxiliary files and implementing them locally, and by downloading the full installation package for the Windows application. The community is encouraged to take advantage of the new models and perform careful comparisons with data.

*Acknowledgments.* We thank Wolf-Rainer Hamann and Helge Todt (Potsdam) for making their W-R spectra available to the community and for help with the download. An anonymous referee provided useful suggestions which helped clarify the presentation. Support for this work has been provided by NASA through grant number AR-12824 from the Space Telescope Science Institute, which is operated by AURA, Inc., under NASA contract NAS5-26555. EML is supported by NASA through Hubble Fellowship grant no. HST-HF-51324.01-A from the Space Telescope Science Institute, which is operated by the Association of Universities for Research in Astronomy, Inc., under NASA contract NAS5-26555.

# Figures

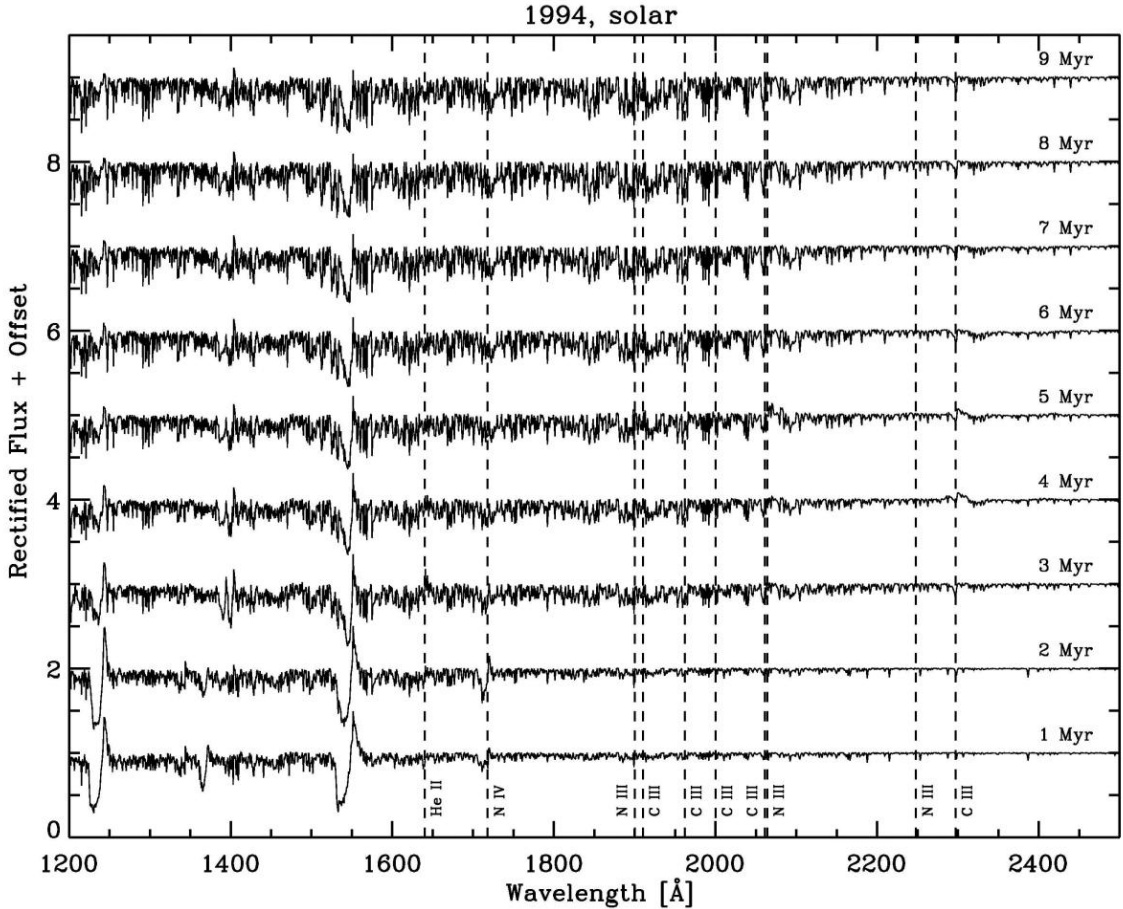

Figure 1. – UV spectral evolution of a single stellar population from 1 to 9 Myr following a Kroupa IMF. Spectral features due to W-R stars are appearing at 3 Myr and later ages. The dashed vertical lines indicate the wavelengths of the most prominent features observed in W-R stars. Most of them are not strong enough in this model but they are present in Figure 5. The strongest W-R features are He II 1640 and N III 2064. These lines are predicted by the W-R models of the newly implemented PoWR library. The 1994-h tracks were used for this sequence.

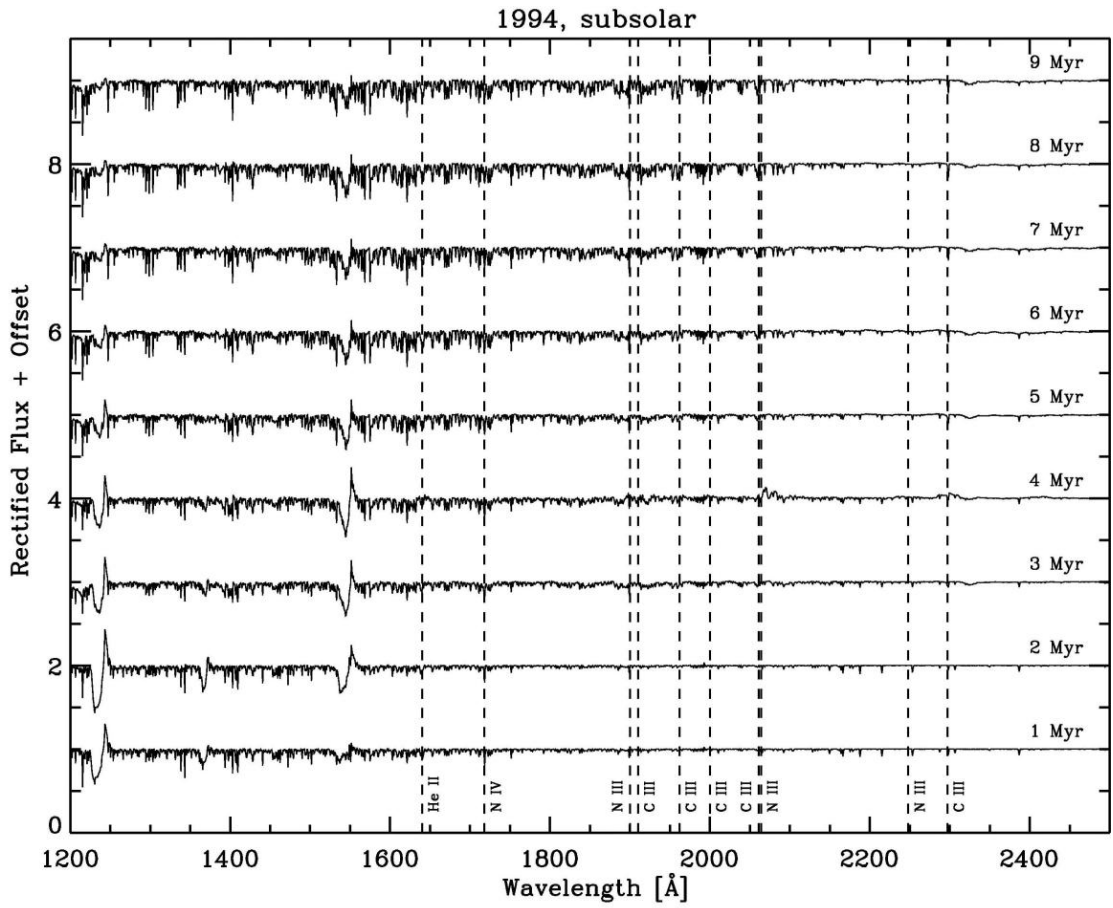

Figure 2. – Same as Figure 1, but for the 1994-l tracks.

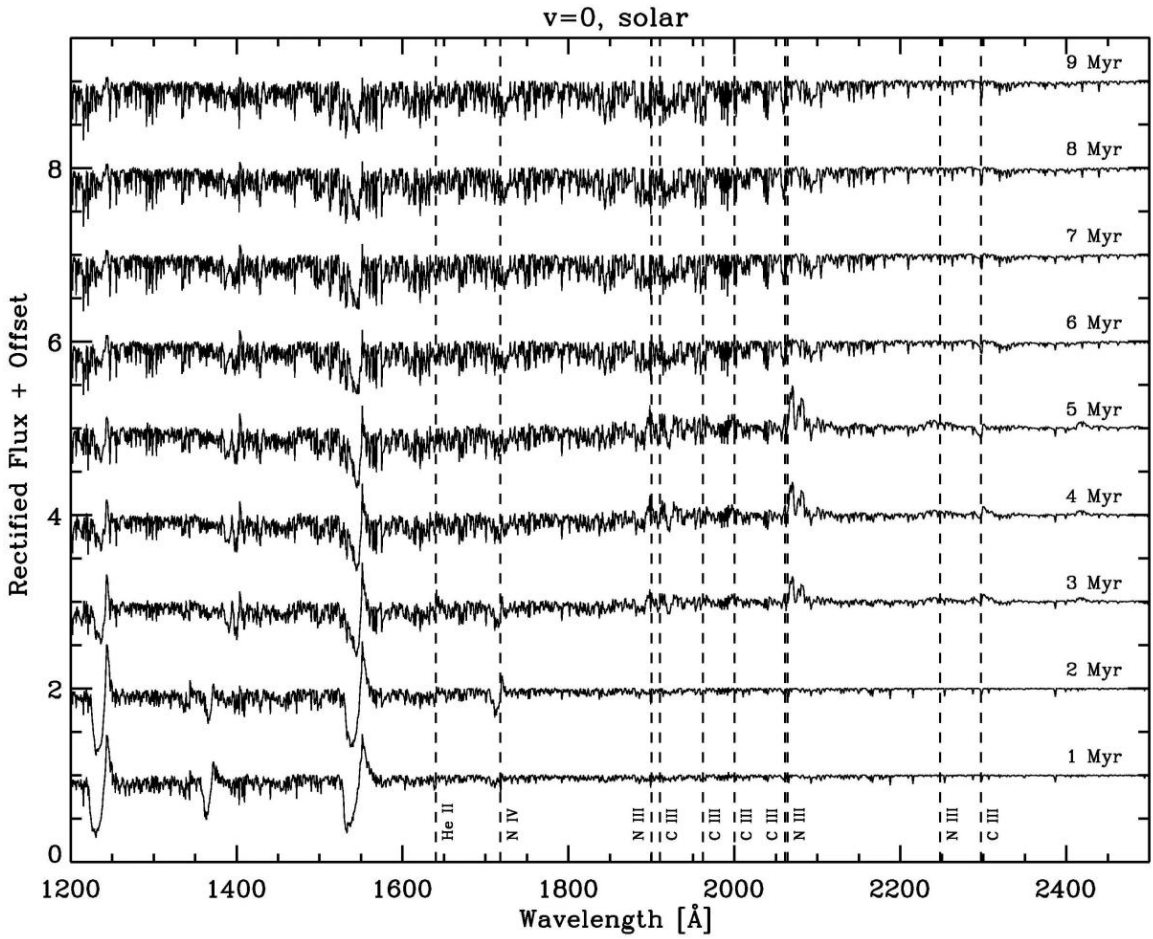

Figure 3. − Same as Figure 1, but for the v00-h evolutionary tracks.

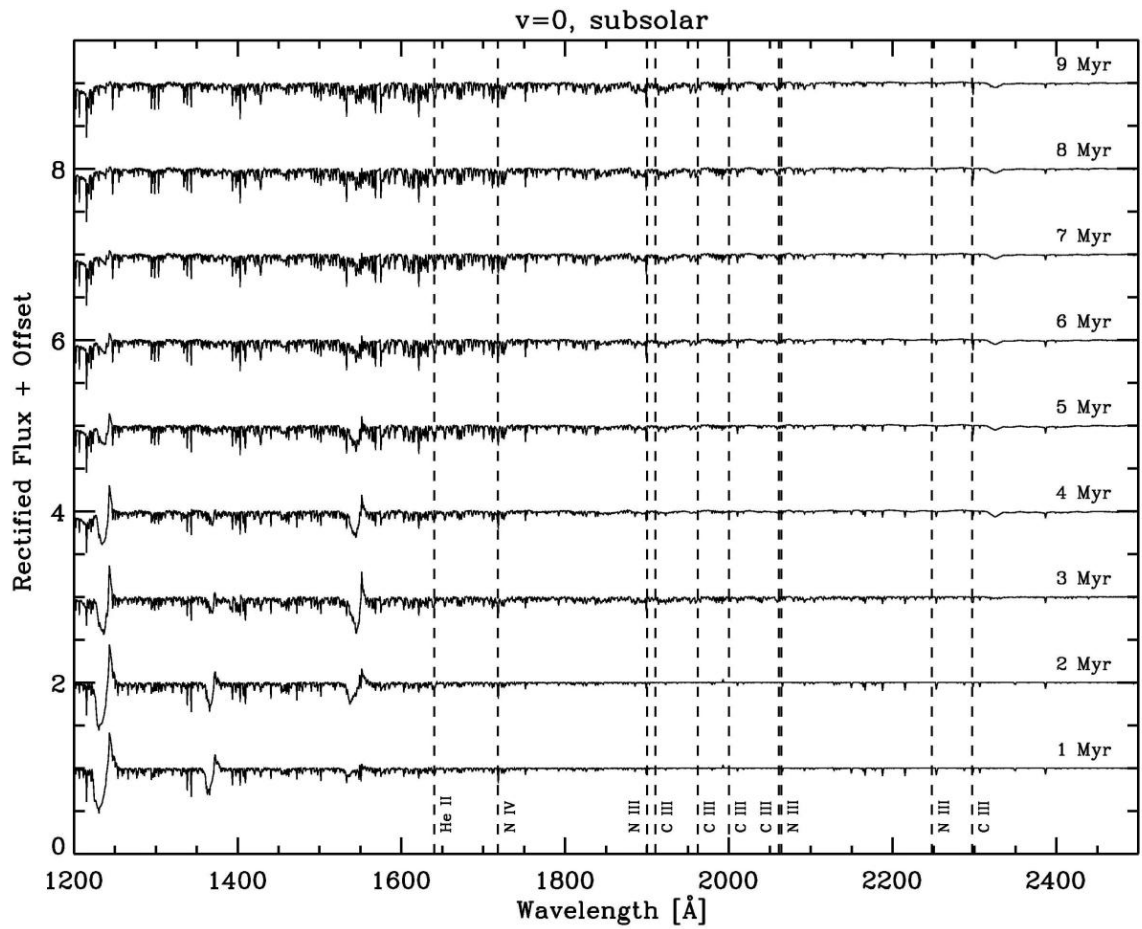

Figure 4. − Same as Figure 3, but for the v00-l evolutionary tracks.

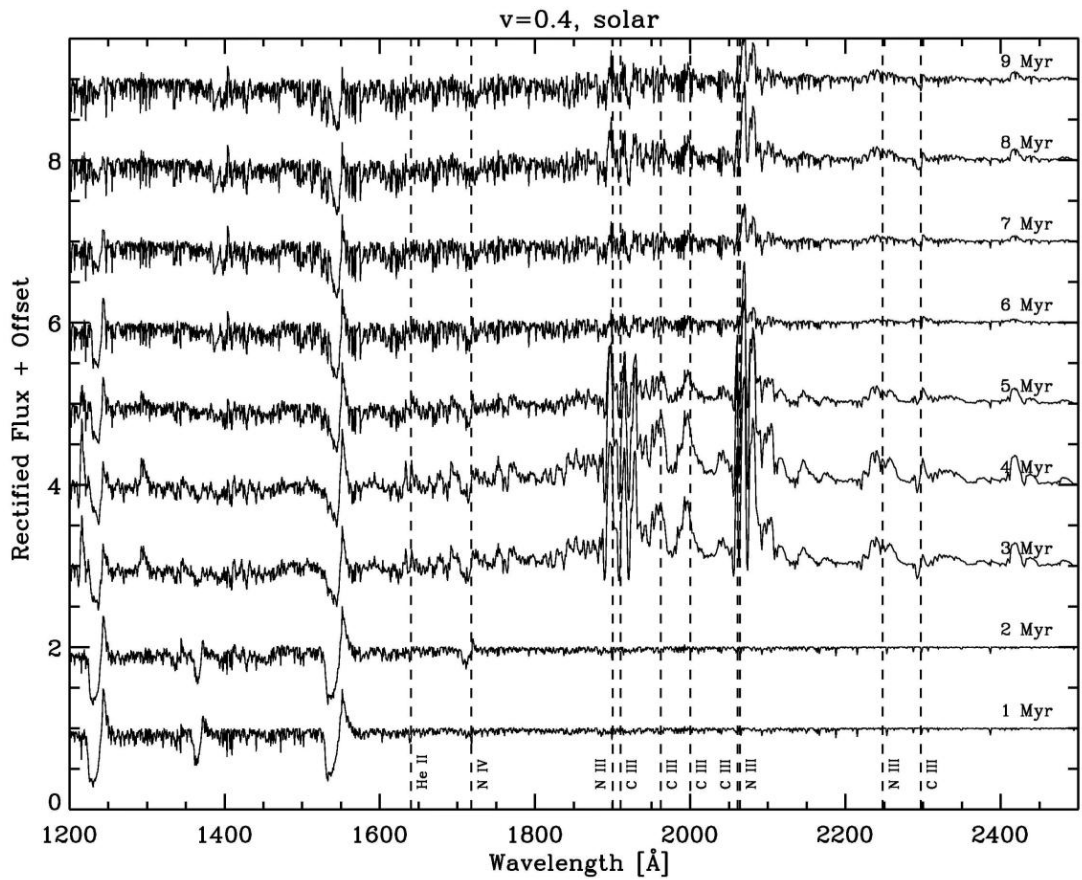

Figure 5. − Same as Figure 1, but for the v40-h evolutionary tracks.

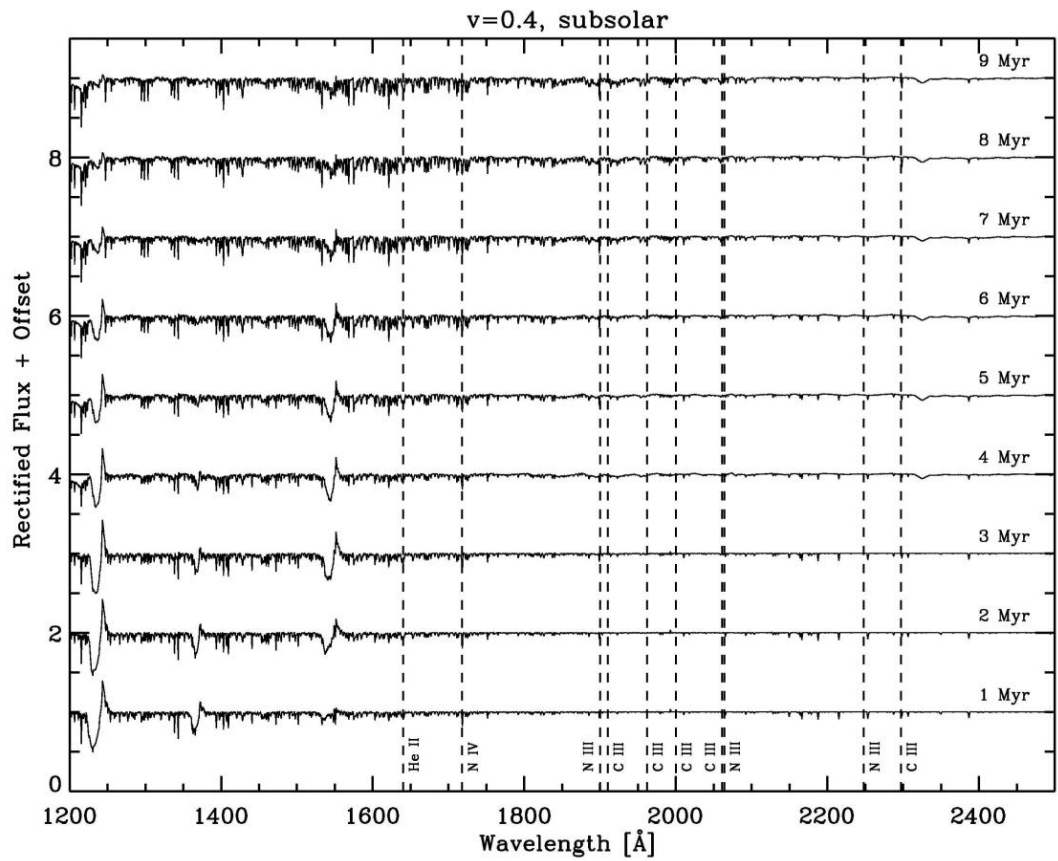

Figure 6. − Same as Figure 5, but for the v40-l evolutionary tracks.

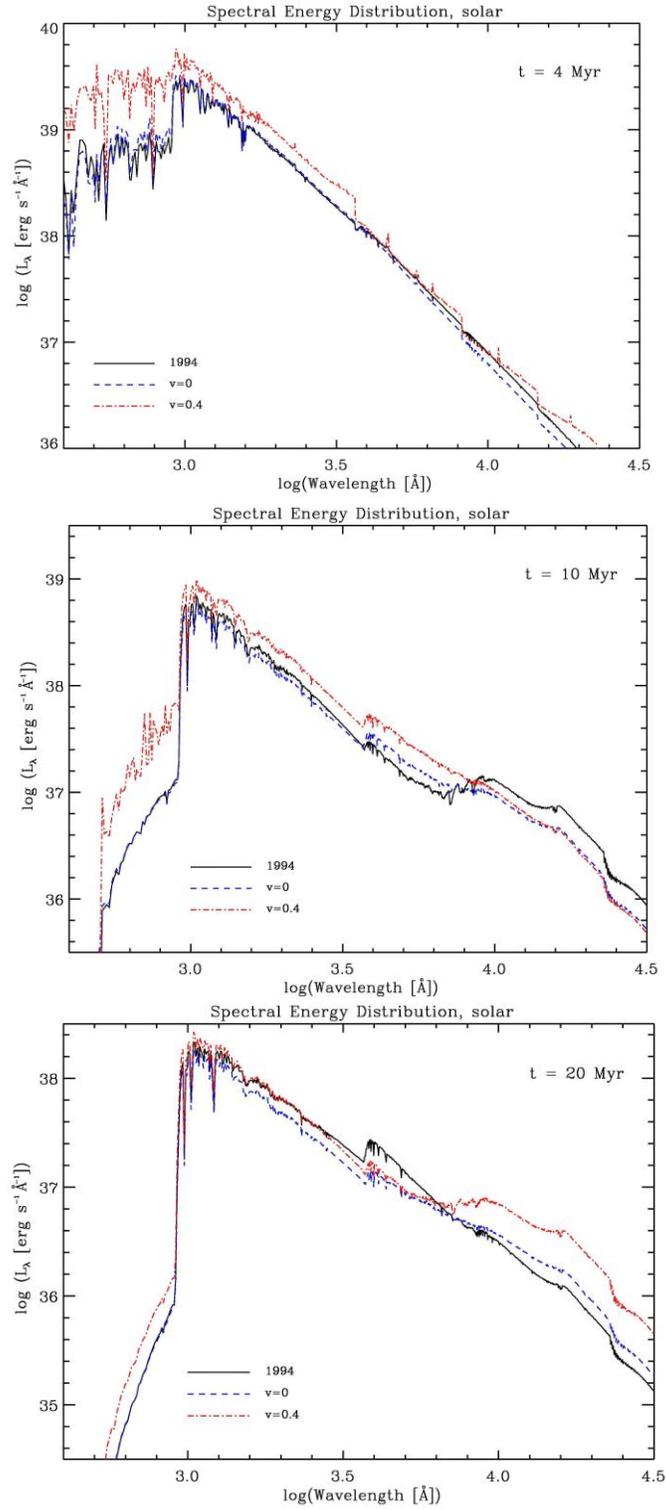

Figure 7. – Comparison of the SSP SEDs at 4 (top), 10 (center), and 20 Myr (bottom) for the 1994-h, v00-h, and v40-h tracks. A Kroupa IMF was used.

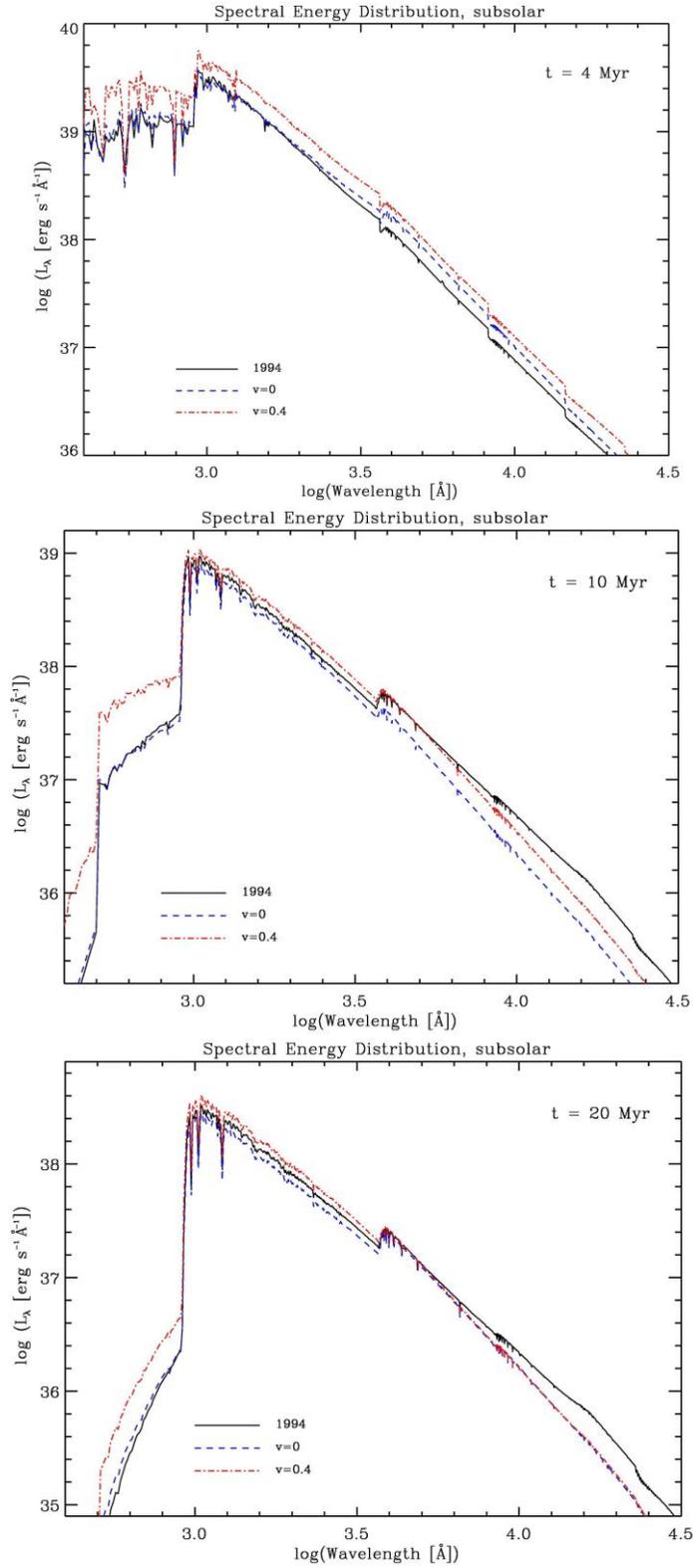

Figure 8. – Same as Figure 7, but for 1994-l, v00-l, and v40-l.

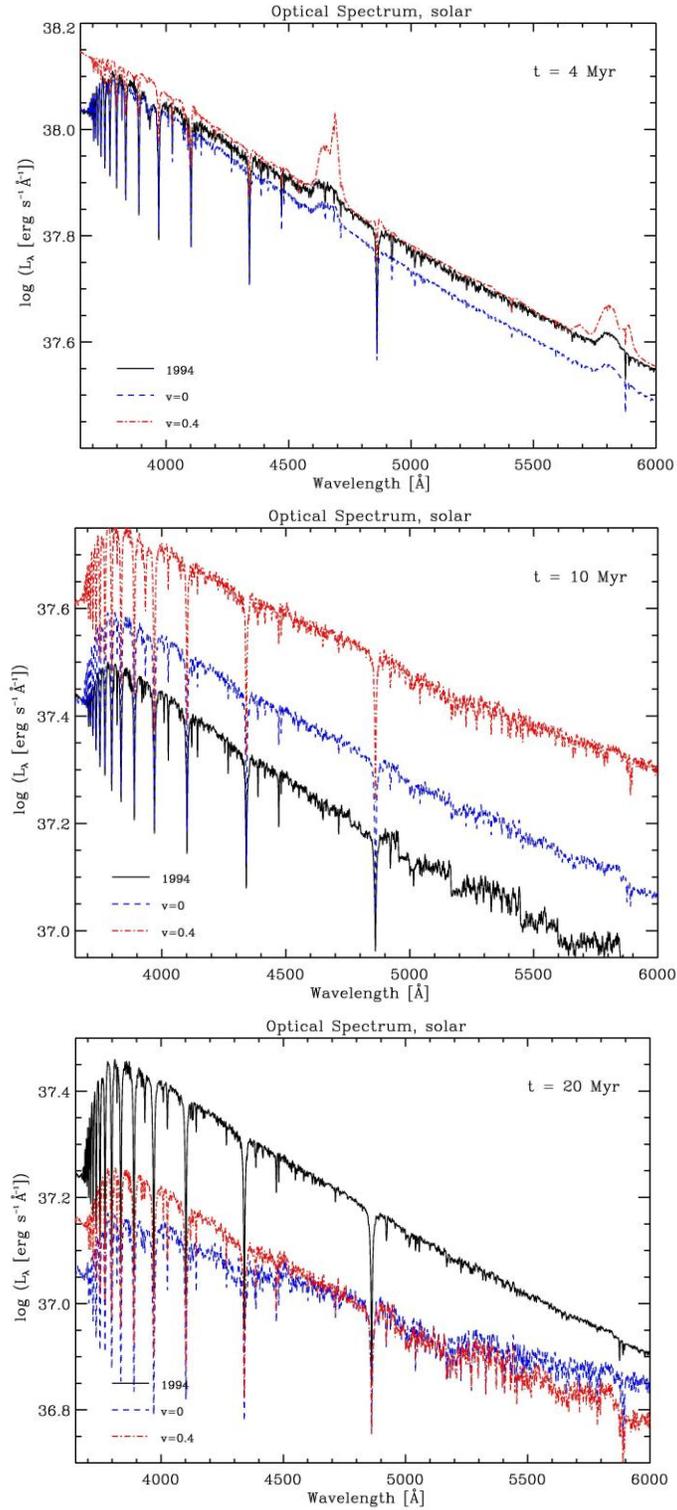

Figure 9. − Comparison of the optical high-resolution spectrum at 4 (top), 10 (center), and 20 Myr (bottom) for the 1994-h, v00-h, and v40-h tracks.

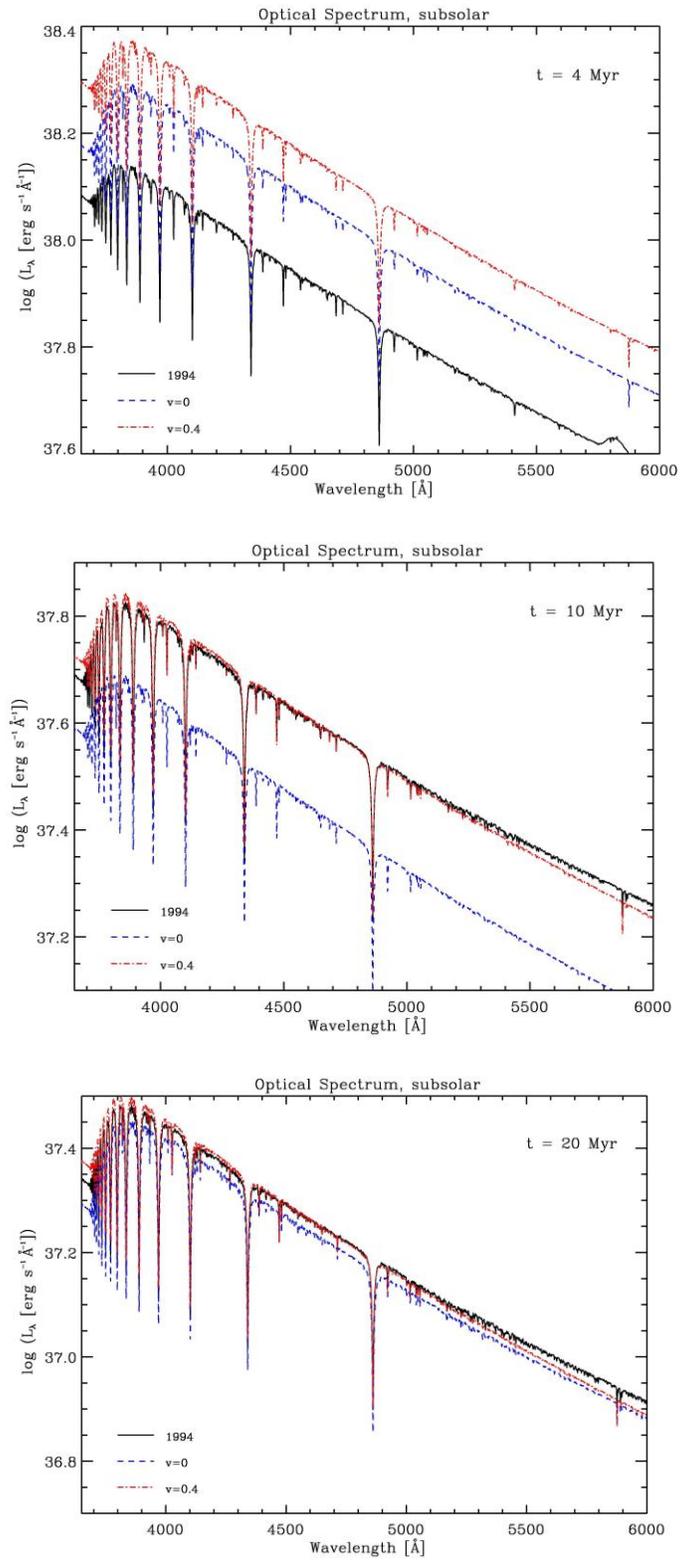

Figure 10. – Same as Figure 9, but for 1994-l, v00-l, and v40-l.

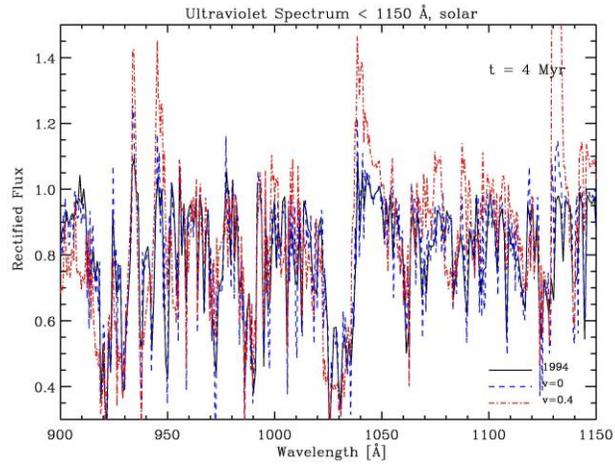
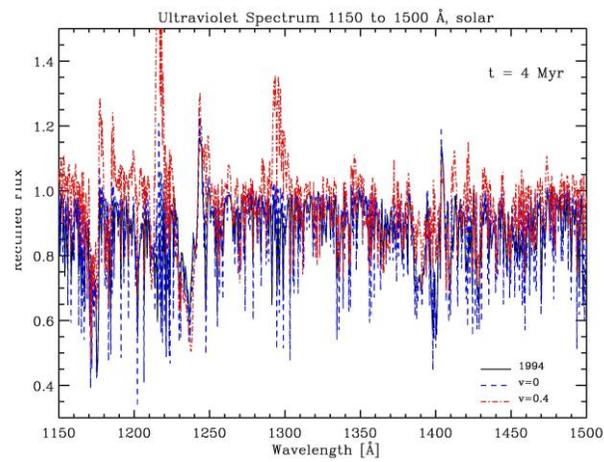
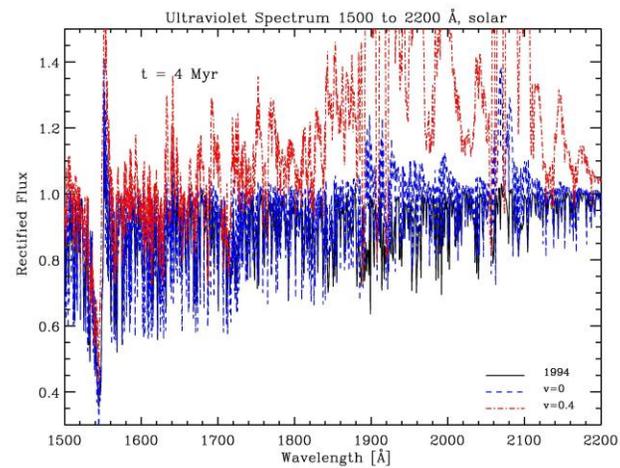

Figure 11. – Comparison of the spectrum at short (top) intermediate (center), and long (bottom) UV wavelengths for the 1994-h, v00-h, and v40-h tracks. The age is 4 Myr.

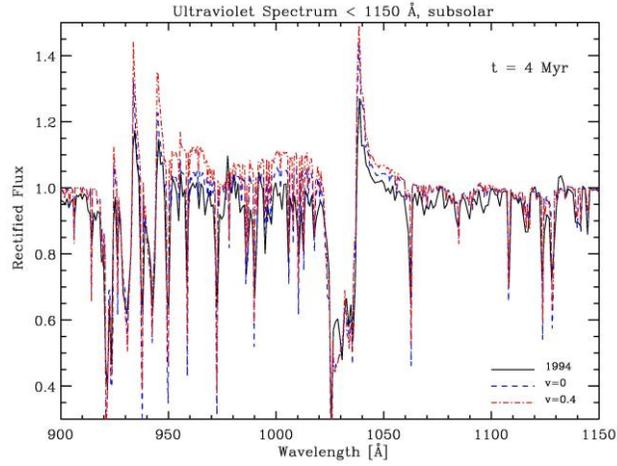
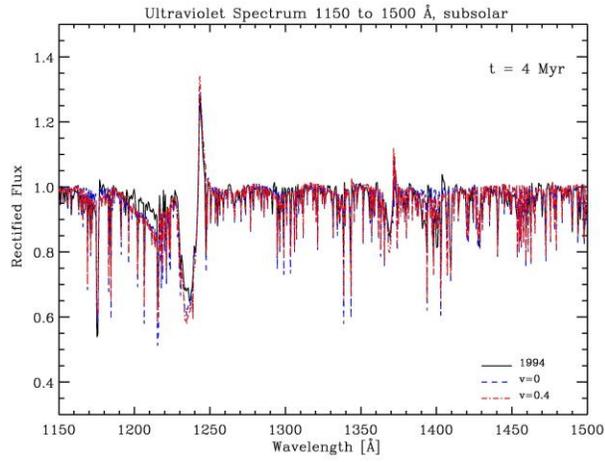
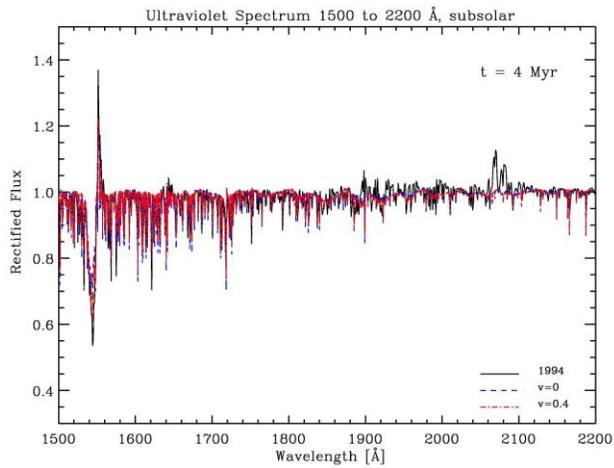

Figure 12. – Same as Figure 11, but for 1994-l, v00-l, and v40-l.

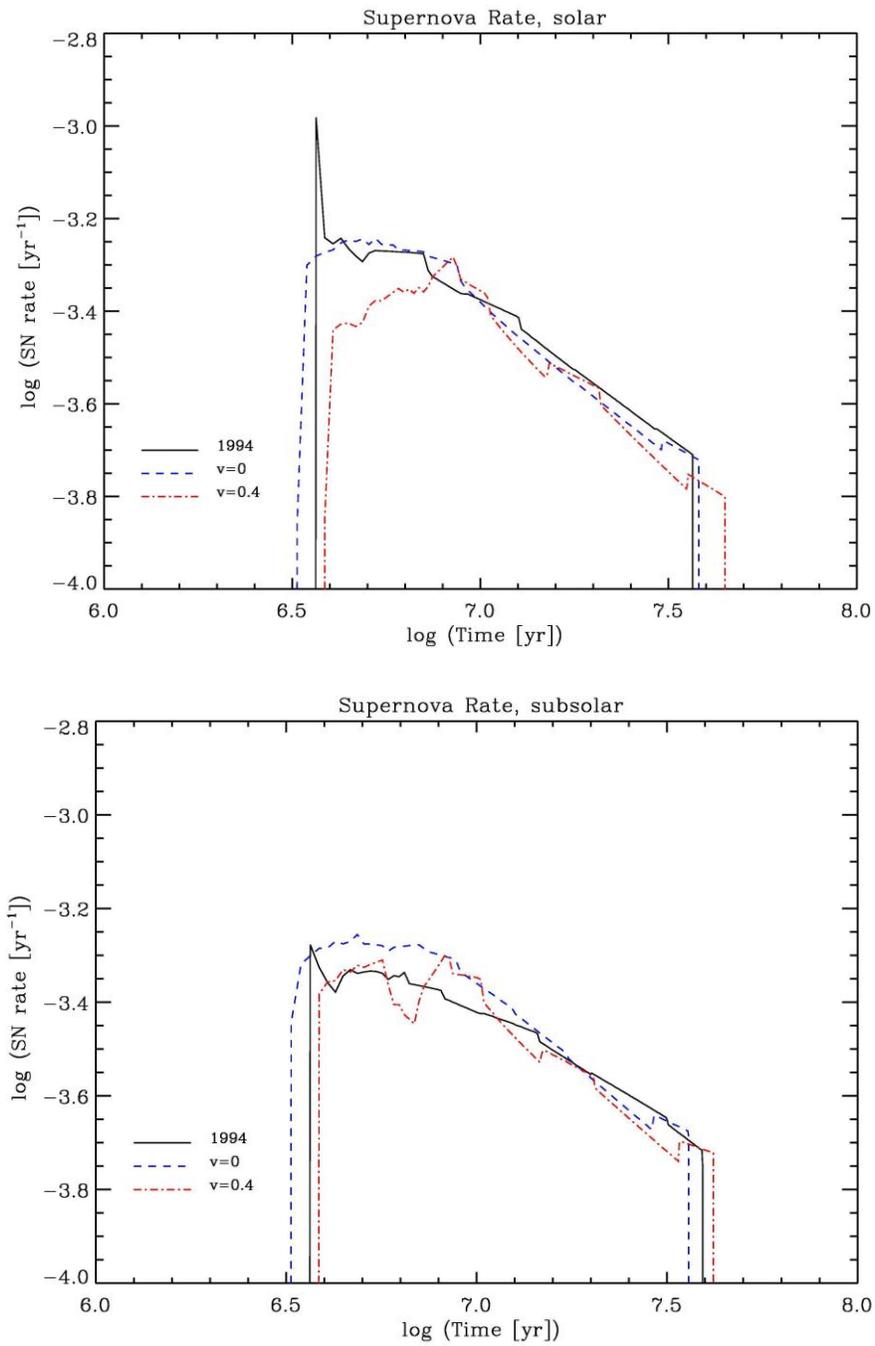

Figure 13. – Core-collapse supernova rates predicted by the 1994, v00, and v40 tracks at solar (top) and subsolar (bottom) chemical composition.

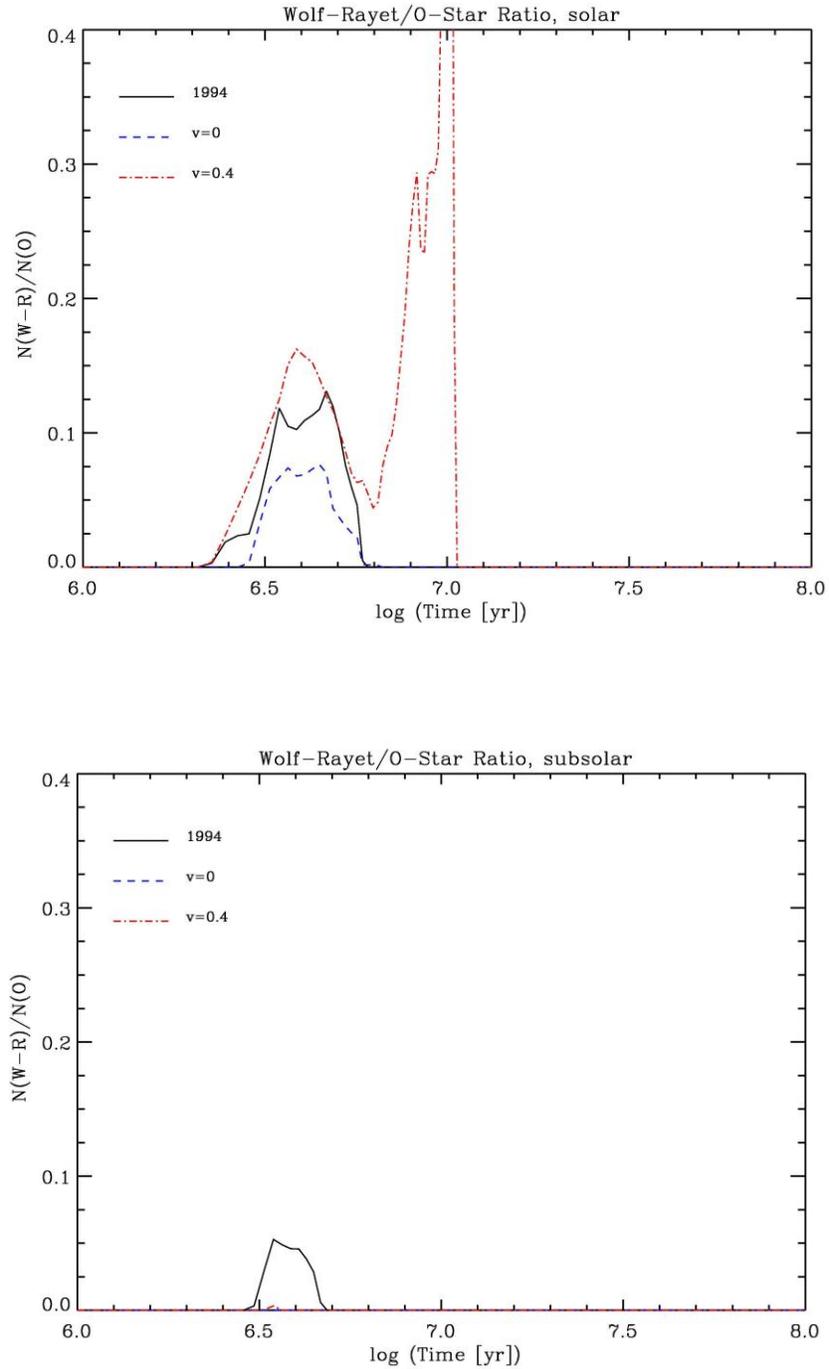

Figure 14. – Ratio of W-R over O stars predicted by the 1994, v00, and v40 tracks at solar (top) and subsolar (bottom) chemical composition. The v00-l and v40-l models produce no significant W-R population.

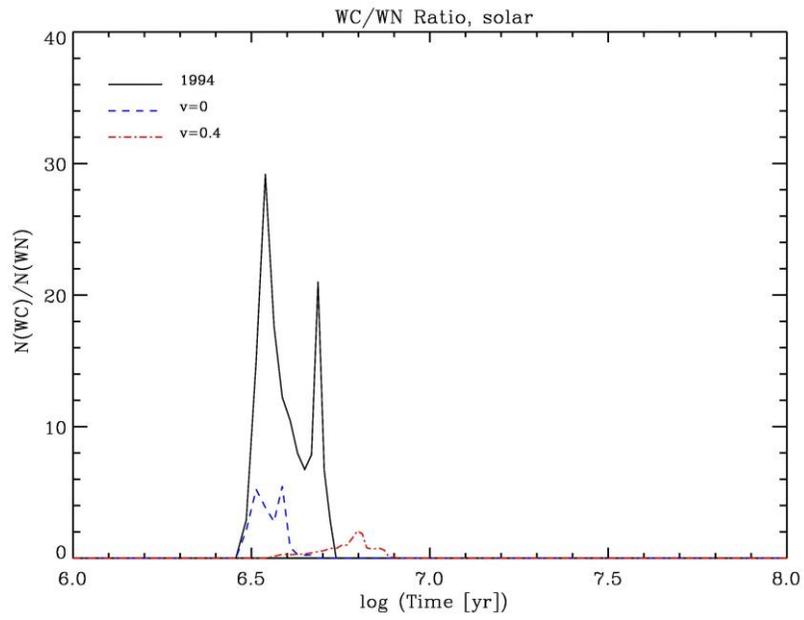

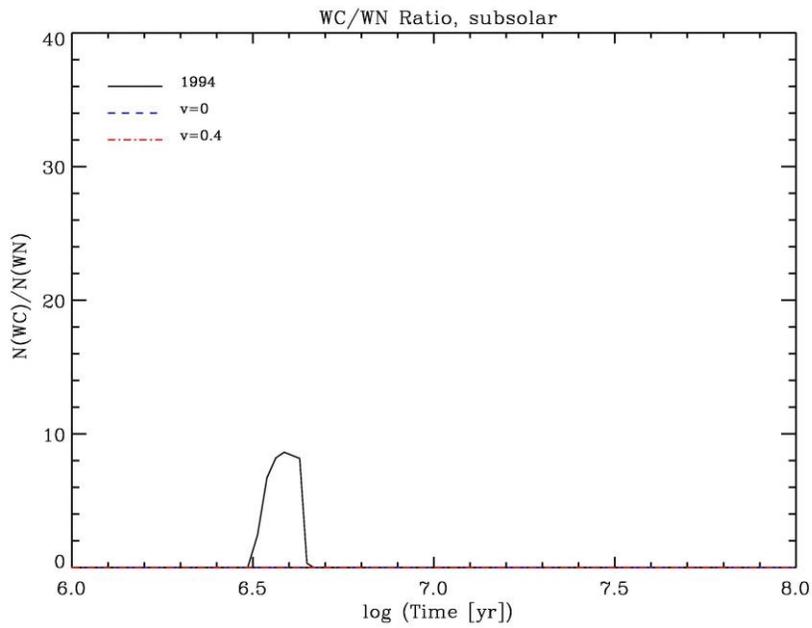

Figure 15. – Ratio of WC over WN stars predicted by the 1994, v00, and v40 tracks at solar (top) and subsolar (bottom) chemical composition.

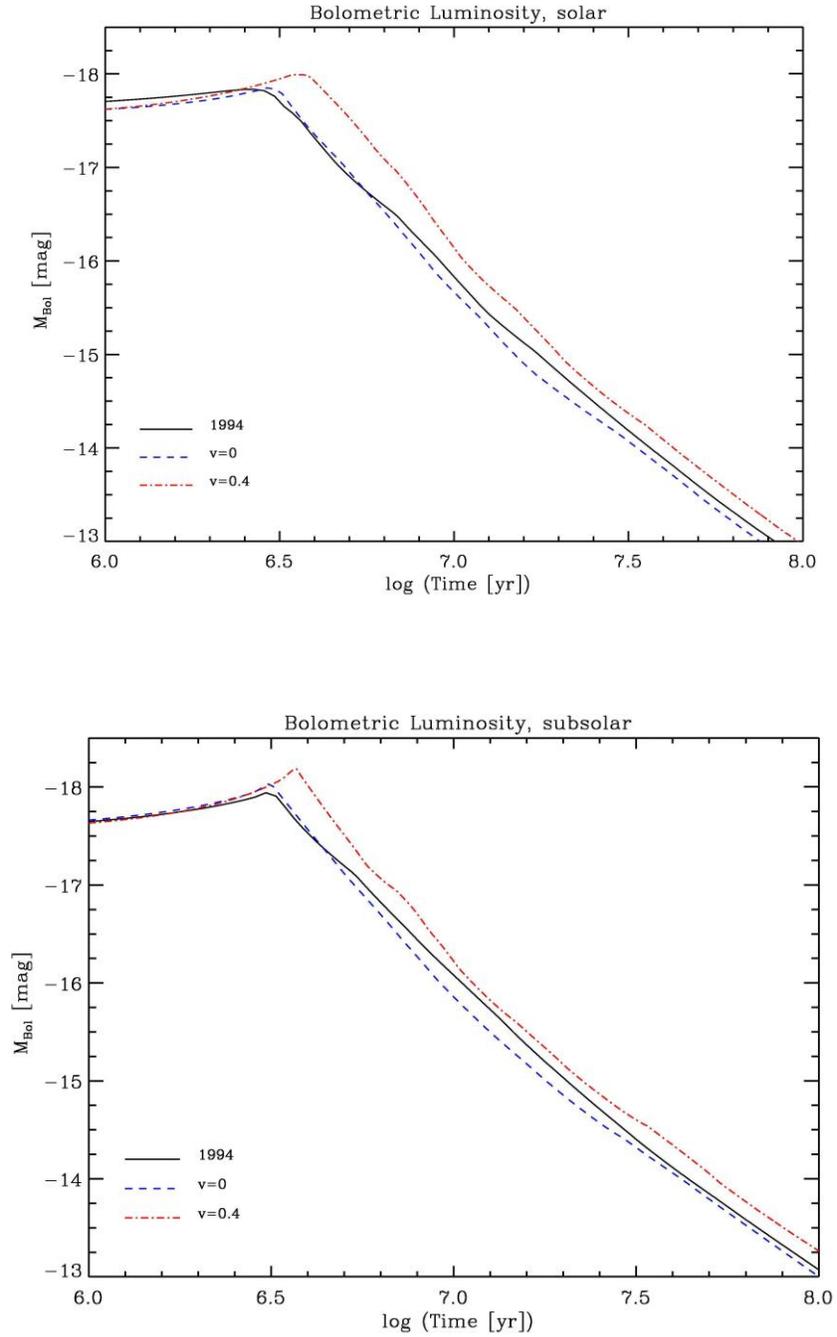

Figure 16. – Bolometric luminosity predicted by the 1994, v00, and v40 tracks at solar (top) and subsolar (bottom) chemical composition.

.

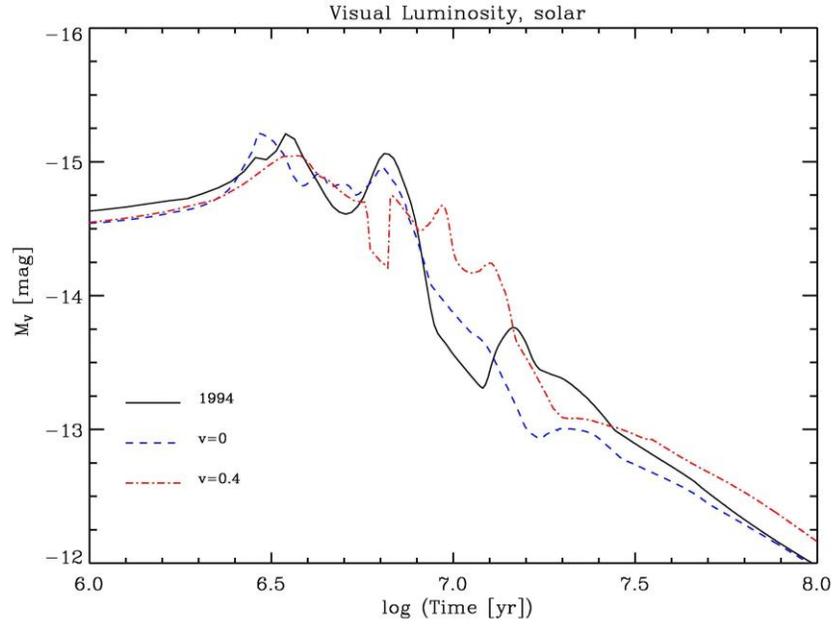

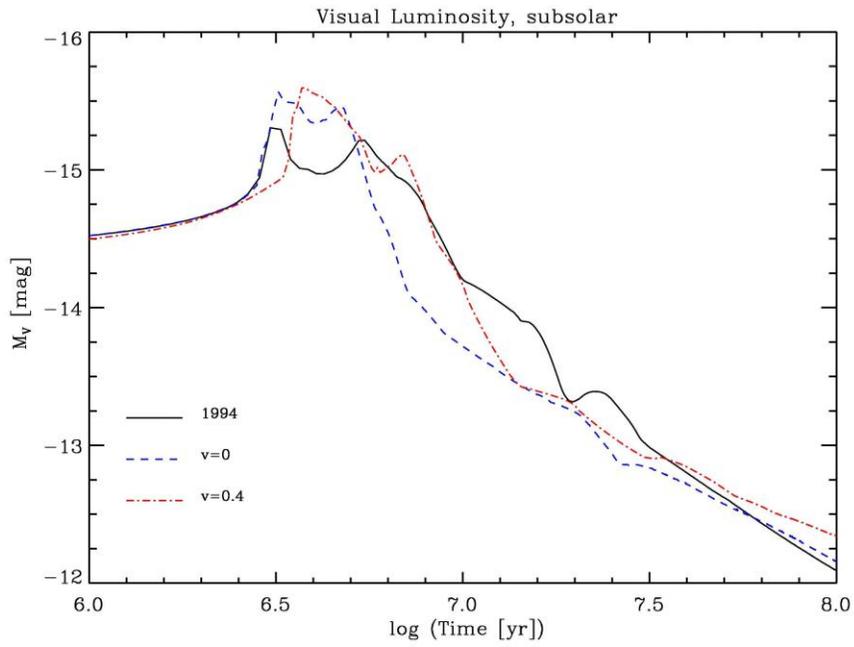

Figure 17. – Visual luminosity predicted by the 1994, v00, and v40 tracks at solar (top) and subsolar (bottom) chemical composition.

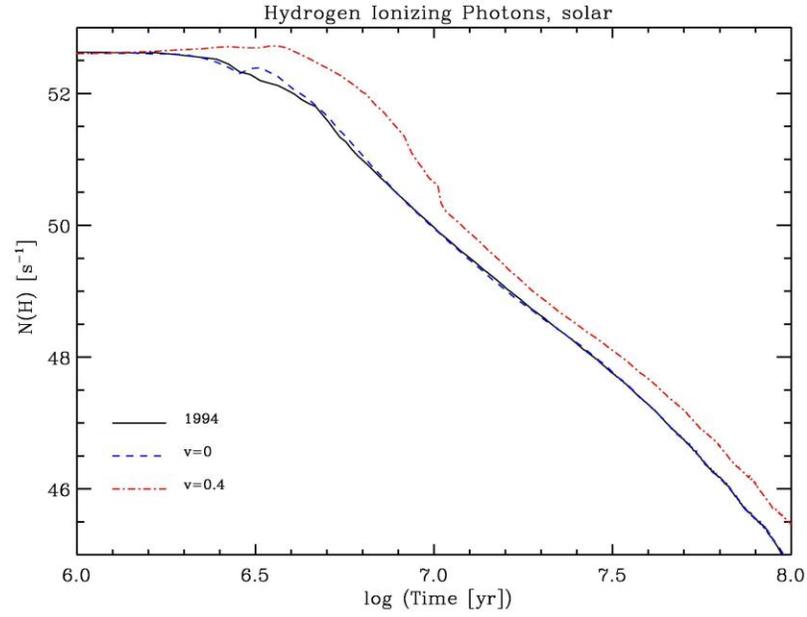

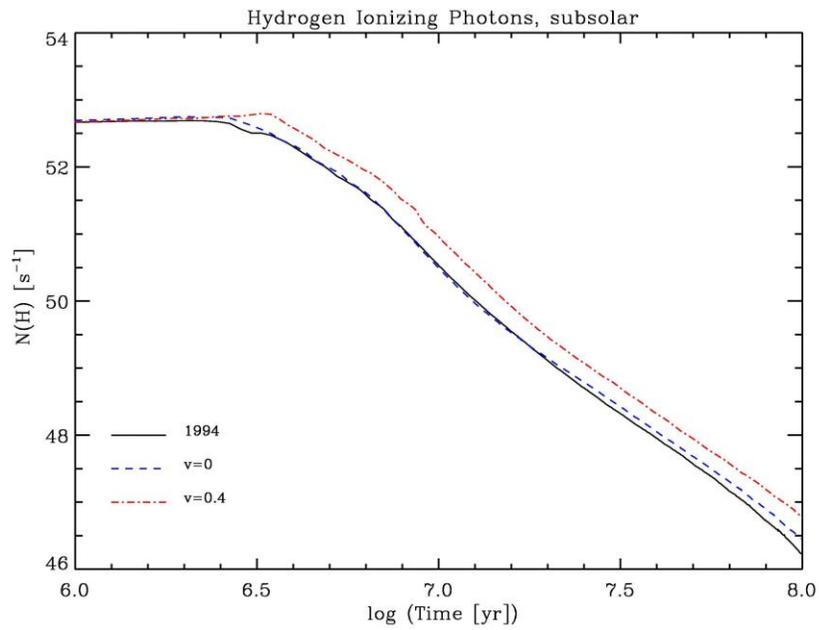

Figure 18. – Number of ionizing photons shortward of 912 Å predicted by the 1994, v00, and v40 tracks at solar (top) and subsolar (bottom) chemical composition.

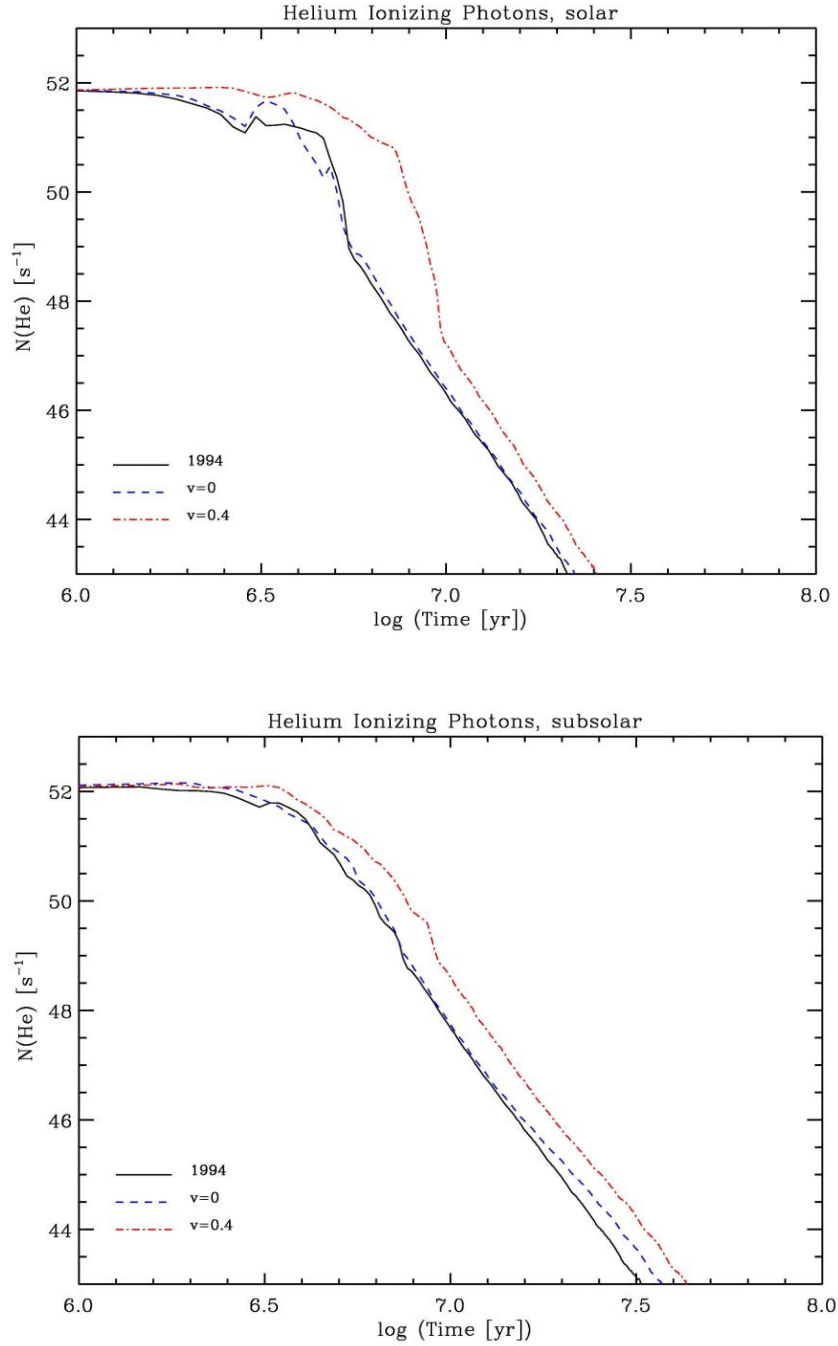

Figure 19. – Number of ionizing photons shortward of 504 Å predicted by the 1994, v00, and v40 tracks at solar (top) and subsolar (bottom) chemical composition.

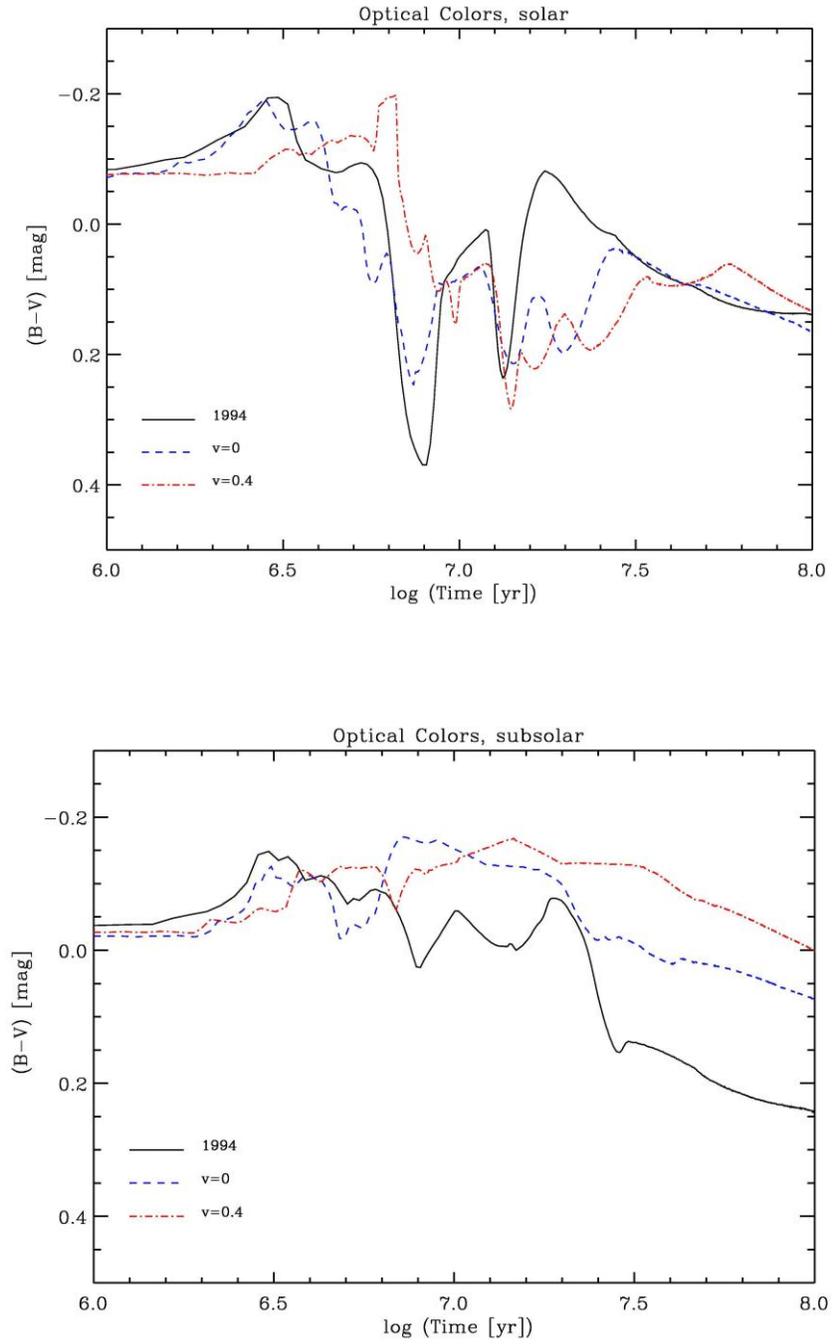

Figure 20. − (*B*−*V*) color predicted by the 1994, v00, and v40 tracks at solar (top) and subsolar (bottom) chemical composition.

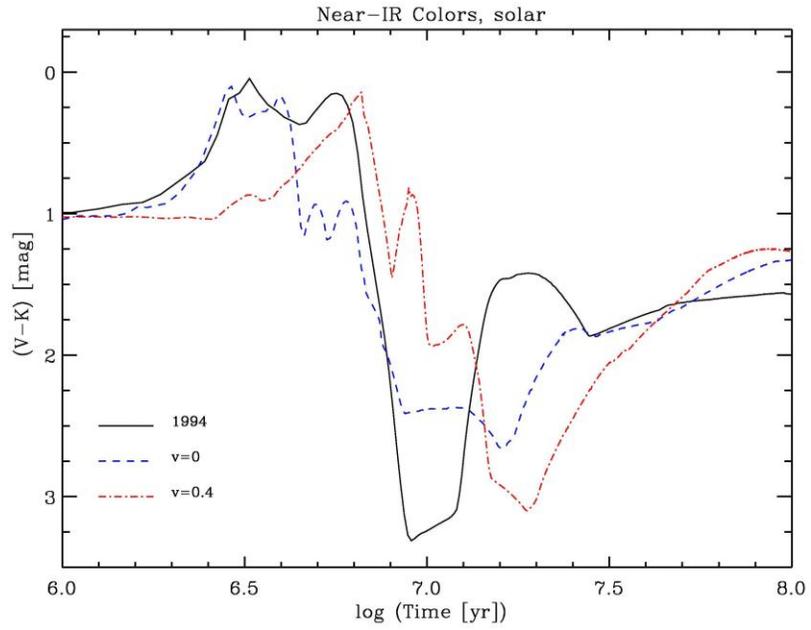

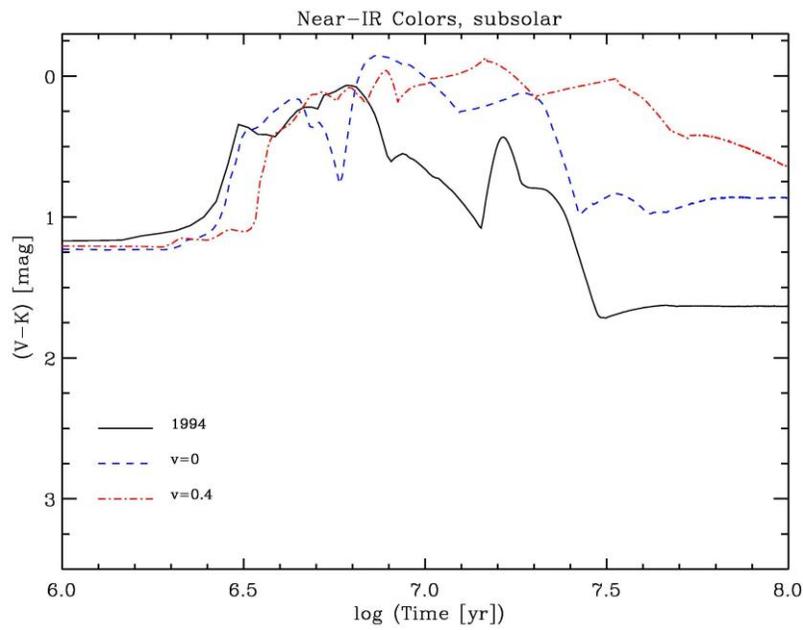

Figure 21. – (*V–K*) color predicted by the 1994, v00, and v40 tracks at solar (top) and subsolar (bottom) chemical composition.

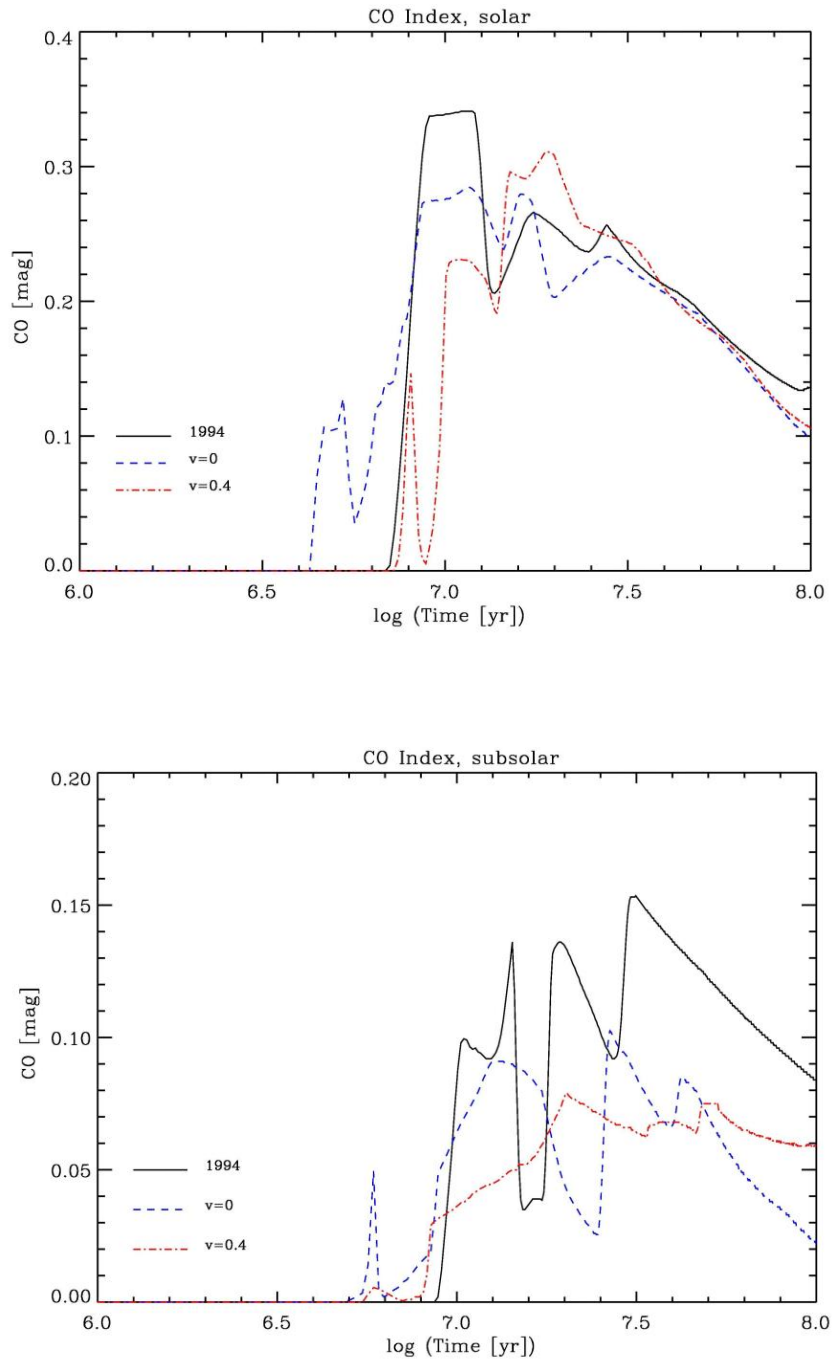

Figure 22. – CO index predicted by the 1994, v00, and v40 tracks at solar (top) and subsolar (bottom) chemical composition.

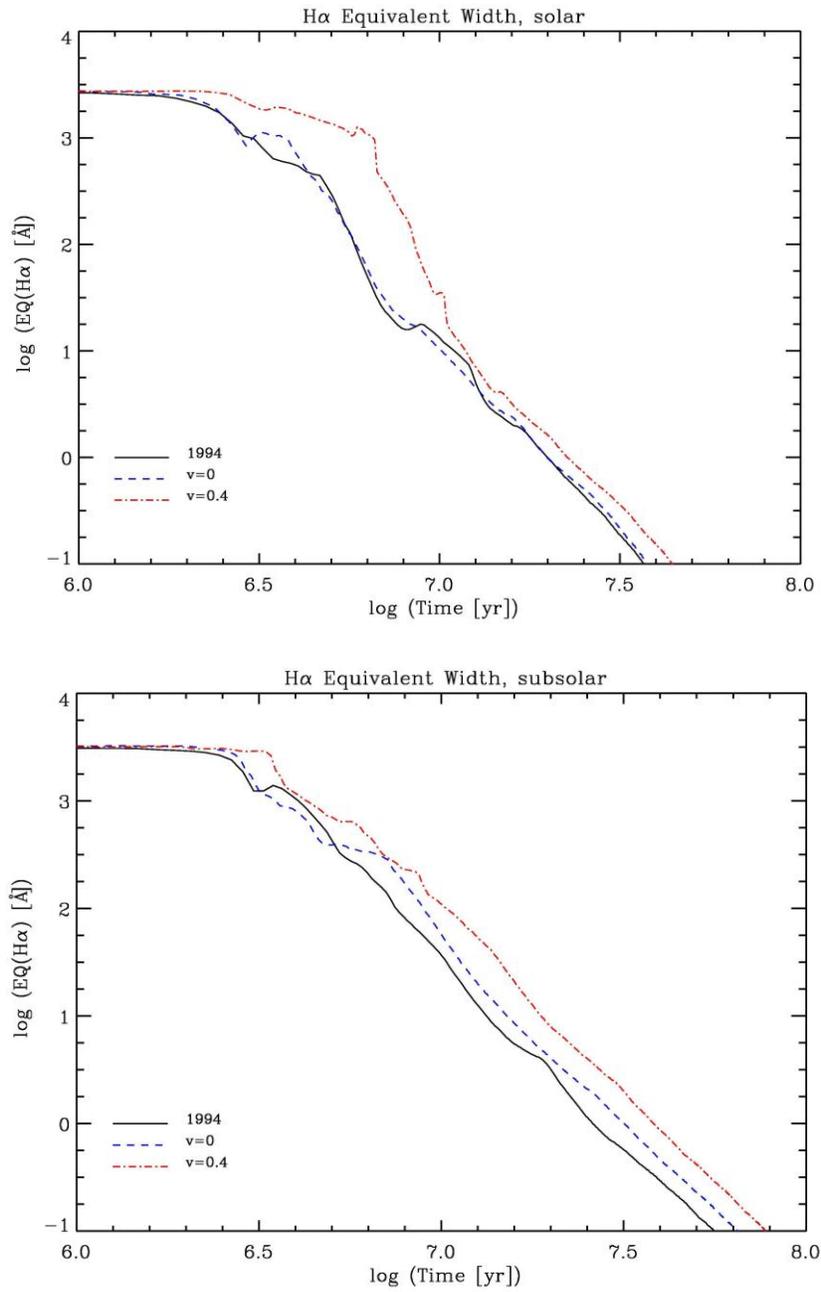

Figure 23. – Hα equivalent width predicted by the 1994, v00, and v40 tracks at solar (top) and subsolar (bottom) chemical composition.

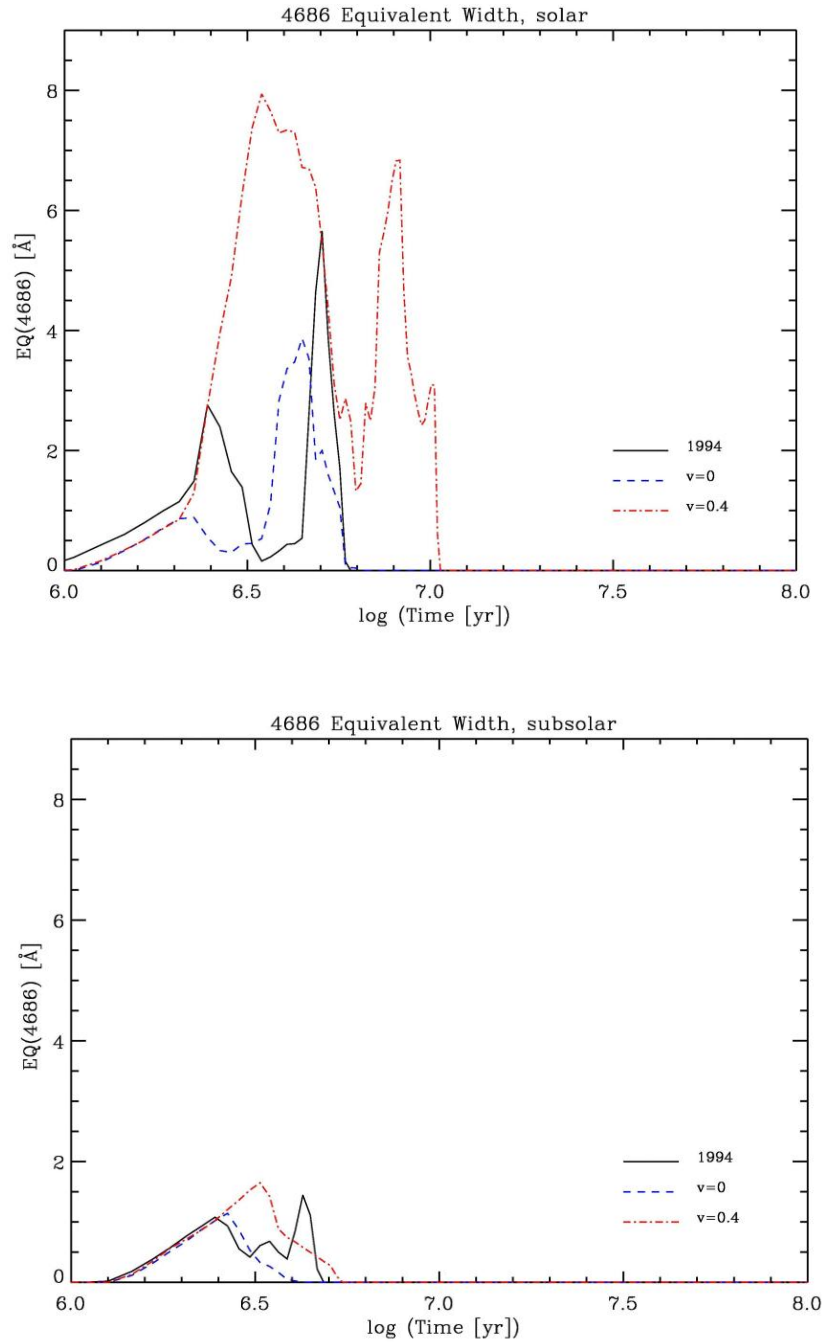

Figure 24. – Equivalent width of *stellar* He II 4686 from W-R stars predicted by the 1994, v00, and v40 tracks at solar (top) and subsolar (bottom) chemical composition.

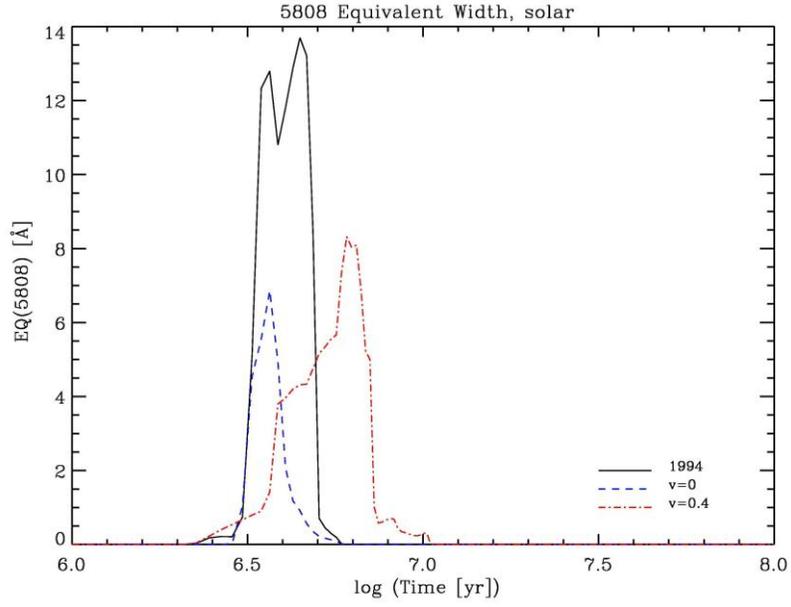
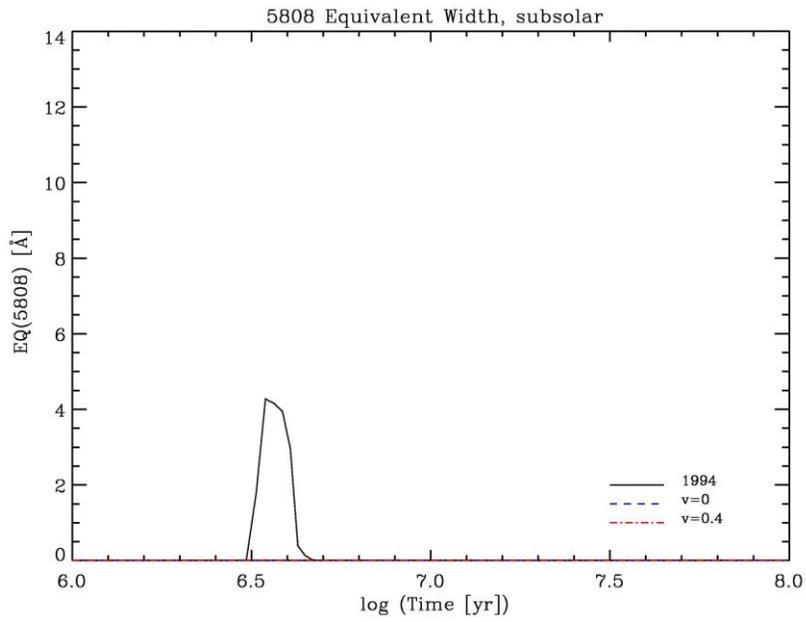

Figure 25. – Equivalent width of *stellar* C IV 5808 from W-R stars predicted by the 1994, v00, and v40 tracks at solar (top) and subsolar (bottom) chemical composition.

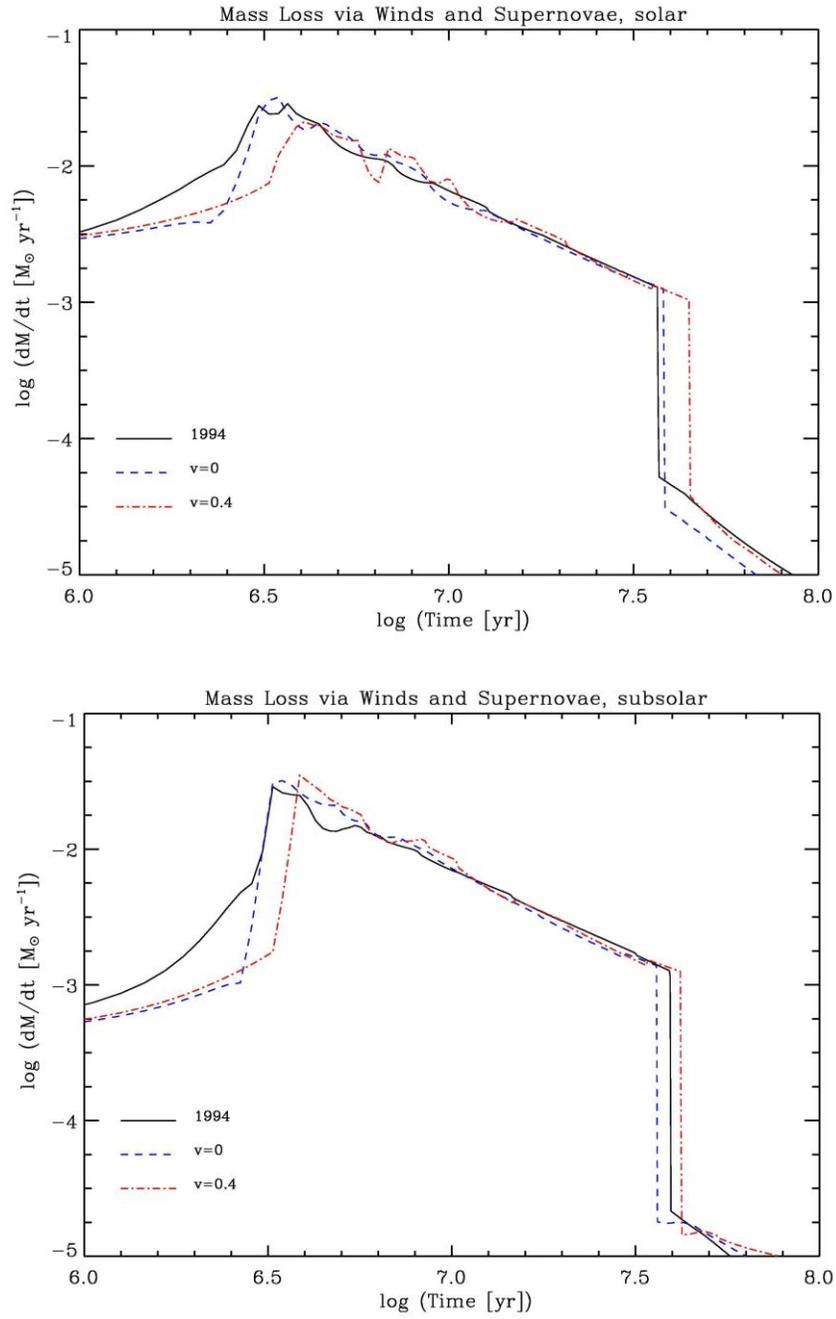

Figure 26. – Mass release by stars and supernovae predicted by the 1994, v00, and v40 tracks at solar (top) and subsolar (bottom) chemical composition.

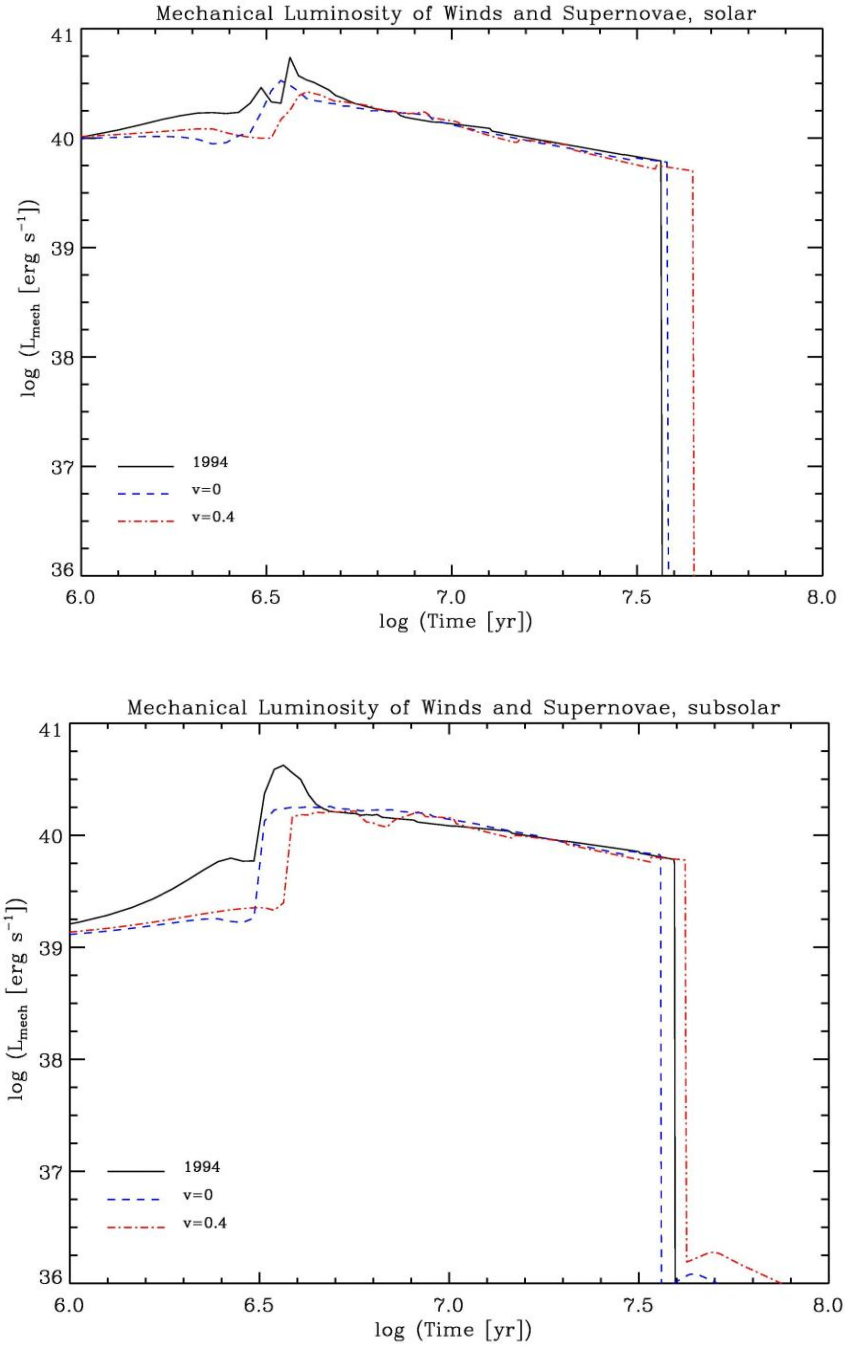

Figure 27. – Mechanical luminosity of stellar winds and supernovae predicted by the 1994, v00, and v40 tracks at solar (top) and subsolar (bottom) chemical composition.

# Tables

Table 1. Properties of the stellar evolution models

| Designation | Reference | $Z$ | $v_{\rm rot}$ (km s$^{-1}$) | $d_{\rm over}/H_{\rm P}$ | $\dot{M}_{\rm O}$ | $\dot{M}_{\rm W\text{-}R}$ | $\dot{M}_{\rm RSG}$ |
|---|---|---|---|---|---|---|---|
| v40-h | E12 | 0.014 | 0.4 $v_{\rm break}$ | 0.1 | (1) | (2), (3) | (4), (5) |
| v40-l | G13 | 0.002 | 0.4 $v_{\rm break}$ | 0.1 | $\propto (Z/Z_\odot)^{0.85-0.5}$ | $\propto (Z/Z_\odot)^{0.66}$ | $\propto (Z/Z_\odot)^{0}$ |
| v00-h | E12 | 0.014 | 0 | 0.1 | (1) | (2), (3) | (4), (5) |
| v00-l | G13 | 0.002 | 0 | 0.1 | $\propto (Z/Z_\odot)^{0.85-0.5}$ | $\propto (Z/Z_\odot)^{0.66}$ | $\propto (Z/Z_\odot)^{0}$ |
| 1994-h | M94 | 0.020 | 0 | 0.2 | 2× (6) | 8×10$^{-5}$ M$_\odot$ yr$^{-1}$, (7) | 2× (6) |
| 1994-l | M94 | 0.004 | 0 | 0.2 | $\propto (Z/Z_\odot)^{0.5}$ | $\propto (Z/Z_\odot)^{0}$ | $\propto (Z/Z_\odot)^{0.5}$ |

Key to mass-loss references:
- (1) Vink et al. (2001)
- (2) Nugis & Lamers (2000)
- (3) Gräfener & Hamann (2008)
- (4) Sylvester et al. (1998)
- (5) van Loon et al. (1999)
- (6) de Jager et al. (1988)
- (7) Langer (1989)

Table 2. Properties of the PoWR library stars

| Label | $\log T_*$ (K) | $R_*$ ($R_\odot$) | $\log \dot{M}$ ($M_\odot\,\mathrm{yr}^{-1}$) | $D$ | $v_\infty$ (km s$^{-1}$) | $\log R_T$ ($R_\odot$) | PoWR Grid # |
|---|---|---|---|---|---|---|---|
| WN1 | 4.50 | 14.92 | −4.8 | 4 | 1000 | 1.24 | WNL 3-9 |
| WN2 | 4.55 | 11.85 | −4.8 | 4 | 1000 | 1.14 | WNL 4-10 |
| WN3 | 4.60 | 9.42 | −4.8 | 4 | 1000 | 1.04 | WNL 5-11 |
| WN4 | 4.65 | 7.48 | −4.8 | 4 | 1000 | 0.94 | WNL 6-12 |
| WN5 | 4.70 | 5.94 | −4.8 | 4 | 2000 | 1.04 | WNE 7-11 |
| WN6 | 4.75 | 4.72 | −4.8 | 4 | 2000 | 0.94 | WNE 8-12 |
| WN7 | 4.80 | 3.75 | −4.8 | 4 | 2000 | 0.84 | WNE 9-13 |
| WN8 | 4.85 | 2.98 | −4.8 | 4 | 2000 | 0.74 | WNE 10-14 |
| WN9 | 4.90 | 2.37 | −4.8 | 4 | 2000 | 0.64 | WNE 11-15 |
| WN10 | 4.95 | 1.88 | −4.8 | 4 | 2000 | 0.54 | WNE 12-16 |
| WN11 | 5.00 | 1.49 | −4.8 | 4 | 2000 | 0.44 | WNE 13-17 |
| WN12 | 5.05 | 1.19 | −4.8 | 4 | 2000 | 0.34 | WNE 14-18 |
| WC1 | 4.65 | 7.48 | −5.0 | 10 | 2000 | 1.14 | WC 6-10 |
| WC2 | 4.70 | 5.94 | −5.0 | 10 | 2000 | 1.04 | WC 7-11 |
| WC3 | 4.75 | 4.72 | −5.0 | 10 | 2000 | 0.94 | WC 8-12 |
| WC4 | 4.80 | 3.75 | −5.0 | 10 | 2000 | 0.84 | WC 9-13 |
| WC5 | 4.85 | 2.98 | −5.0 | 10 | 2000 | 0.74 | WC 10-14 |
| WC6 | 4.90 | 2.37 | −5.0 | 10 | 2000 | 0.64 | WC 11-15 |
| WC7 | 4.95 | 1.88 | −5.0 | 10 | 2000 | 0.54 | WC 12-16 |
| WC8 | 5.00 | 1.49 | −5.0 | 10 | 2000 | 0.44 | WC 13-17 |
| WC9 | 5.05 | 1.19 | −5.0 | 10 | 2000 | 0.34 | WC 14-18 |
| WC10 | 5.10 | 0.94 | −5.0 | 10 | 2000 | 0.24 | WC 15-19 |
| WC11 | 5.15 | 0.75 | −5.0 | 10 | 2000 | 0.14 | WC 16-20 |
| WC12 | 5.20 | 0.59 | −5.0 | 10 | 2000 | 0.04 | WC 17-21 |